\documentclass[namedreferences]{solarphysics}
\usepackage[hyperref,optionalrh,solaromanenum]{spr-sola-addons} 
\usepackage{graphicx}                    
\usepackage{courier}                    
\usepackage{color}                       
\usepackage{breakurl}                         
\usepackage{aas_macros}
\usepackage[all]{hypcap}
\usepackage{upgreek}

\newcommand{\biborder}[1]{}

\begin{document}

\begin{article}

\begin{opening}

\title{Solar X-ray Monitor On Board the Chandrayaan-2 Orbiter: In-flight Performance and Science Prospects}

%
\author[addressref={1,2},corref,email={mithun@prl.res.in}]{\inits{}\fnm{N. P. S. }\lnm{Mithun}\orcid{0000-0003-3431-6110}}
\author[addressref={1}]{\inits{}\fnm{Santosh V. }\lnm{Vadawale}\orcid{0000-0002-2050-0913}}
\author[addressref={1}]{\inits{}\fnm{Aveek }\lnm{Sarkar}\orcid{0000-0002-4781-5798}}
\author[addressref={1}]{\inits{}\fnm{M. }\lnm{Shanmugam}}
\author[addressref={1}]{\inits{}\fnm{Arpit R. }\lnm{Patel}\orcid{0000-0002-0929-1401}}
\author[addressref={1}]{\inits{}\fnm{Biswajit }\lnm{Mondal}}
\author[addressref={1}]{\inits{}\fnm{Bhuwan }\lnm{Joshi}\orcid{0000-0001-5042-2170}}
\author[addressref={1}]{\inits{}\fnm{Janardhan }\lnm{P.}\orcid{0000-0003-2504-2576}}
\author[addressref={1}]{\inits{}\fnm{Hiteshkumar L. }\lnm{Adalja}}
\author[addressref={1}]{\inits{}\fnm{Shiv Kumar }\lnm{Goyal}\orcid{0000-0002-3153-537X}}
\author[addressref={1}]{\inits{}\fnm{Tinkal }\lnm{Ladiya}}
\author[addressref={1}]{\inits{}\fnm{Neeraj Kumar }\lnm{Tiwari}}
\author[addressref={1}]{\inits{}\fnm{Nishant }\lnm{Singh}}
\author[addressref={1}]{\inits{}\fnm{Sushil }\lnm{Kumar}}
\author[addressref={3}]{\inits{}\fnm{Manoj K. }\lnm{Tiwari}\orcid{0000-0001-5143-1423}}
\author[addressref={3}]{\inits{}\fnm{M. H. }\lnm{Modi}}
\author[addressref={1}]{\inits{}\fnm{Anil }\lnm{Bhardwaj}\orcid{0000-0003-1693-453X}}

%
\runningauthor{N. P. S. Mithun et al.}
\runningtitle{Chandrayaan-2 XSM: In-flight Performance and Prospects}

\address[id={1}]{Physical Research Laboratory, Navrangpura, Ahmedabad, 380 009, India}
\address[id={2}]{Indian Institute of Technology Gandhinagar, Palaj, Gandhinagar, 382 355, India}
\address[id={3}]{Raja Ramanna Centre for Advanced Technology, Indore, 452 013, India}

\begin{abstract}
The \textit{Solar X-ray Monitor} (abbreviated as XSM) on board India's
Chandrayaan-2 mission is designed to carry out broadband spectroscopy of
the Sun from lunar orbit. It observes the Sun as a star and measures the
spectrum every second in the soft X-ray band of 1 -- 15 $\mathrm{keV}$ with an energy
resolution better than 180 $\mathrm{eV}$ at 5.9 $\mathrm{keV}$. 
The primary objective of the XSM is to provide the incident solar spectrum for 
the X-ray fluorescence spectroscopy experiment on the Chandrayaan-2 orbiter, which aims to generate 
elemental abundance maps of the lunar surface. However, observations with the XSM can
independently be used to study the Sun as well.
The Chandrayaan-2 mission was launched on 22 July 2019, and the XSM began
nominal operations, in lunar orbit, from September 2019. The in-flight
observations, so far, have shown that its spectral performance has been identical
to that on the ground. Measurements of the effective area from ground calibration
were found to require some refinement, which has been carried out using
solar observations at different incident angles. It also has been shown that
the XSM is sensitive enough to detect solar activity well below A-class.
This makes the investigations of microflares and the quiet solar corona
feasible in addition to the study of the evolution of physical parameters during
intense flares. This article presents the in-flight performance and
calibration of the XSM instrument and discusses some specific science cases
that can be addressed using observations with the XSM.
\end{abstract}

%

\end{opening}

%

\section {Introduction}

The \textit{Solar X-ray Monitor} (XSM) is an X-ray spectrometer on board the 
orbiter of Chandrayaan-2, the second Indian mission to the Moon~\citep{vanitha20}, 
and is a part of the remote X-ray fluorescence spectroscopy experiment of the mission 
aimed at estimating abundances on the lunar surface at a global scale. 
The remote X-ray fluorescence spectroscopy technique has been employed in several 
missions to various atmosphere-less solar system bodies to determine the elemental composition of 
their surfaces~\citep{2007P&SS...55.1135B,bhardwaj_14}. It involves measurement of the characteristic 
X-ray fluorescence lines of various elements emitted due to their 
with the incident solar X-rays. It is possible to obtain a quantitative estimate of 
the abundances of major elements using this technique if the incident X-ray spectrum 
and the observation geometry are known. Such experiments are typically carried out 
from an orbital platform, and hence the information on the observation geometry is 
available. On the other hand, it is known that the X-rays from the Sun are highly variable 
in both their intensity as well as spectral shape. Thus, it is customary to have a separate instrument 
on the same orbital platform that provides simultaneous measurements of the solar X-ray 
spectrum in order to obtain quantitative estimates of the surface elemental abundances. 
On the Chandrayaan-2 mission, the \textit{Chandrayaan-2 Large Area Soft X-ray Spectrometer} 
(CLASS)~\citep{radhakrishna20} and the \textit{Solar X-ray Monitor} (XSM)~\citep{shanmugam20} 
instruments record, respectively, the fluorescence X-ray spectrum from the lunar surface and 
the solar X-ray spectrum.

Remote X-ray fluorescence spectroscopy experiments have been carried out on
several past missions to various solar system objects, such as the Moon
(Apollo-15 and 16, Smart-1, Chandrayaan-1, Chang'e-2, Kaguya), Mercury (Messenger, BeppiColombo), and 
asteroids (NEAR, OSIRIS-REx). All these missions carried dedicated 
instruments for spatially integrated and spectrally resolved solar observations: 
X-ray Solar Monitors on board SMART-1~\citep{2002P&SS...50.1345H}, 
Chandrayaan-1~\citep{2009NIMPA.607..544A}, and Chang'e-2~\citep{2019SoPh..294..120D}; 
MESSENGER-SAX~\citep{2007SSRv..131..393S}; Beppicolombo-SIXS~\citep{2010P&SS...58...96H}; 
and Solar X-ray Monitors on board NEAR-Shoemaker~\citep{2001M&PS...36.1605T} and 
OSIRIS-REx~\citep{2018SSRv..214...48M}. 
Although these instruments' primary objective was to aid the measurement of 
elemental abundances on the surface of the solar system objects, observations with many of them have also been used
for carrying out independent solar studies~\citep{2014SoPh..289.1585N,2015ApJ...803...67D}.

Spectroscopic observations of the Sun in X-ray wavelengths have
contributed enormously to our present understanding of the fundamental
parameters of the solar corona~\citep{2018LRSP...15....5D}. Such studies typically make use 
of the solar X-ray instruments that fall into two classes viz. X-ray imagers providing 
high spatial resolution images of the Sun over a broad energy range, 
but without or with limited spectral information (e.g., Hinode XRT) 
and crystal spectrometers that provide very high-resolution spectra without or with little spatial information, 
but over a narrow energy band (e.g., CORONAS-F RESIK). 
Exceptions to these two categories include RHESSI~\citep{2002SoPh..210....3L} that 
performed imaging spectroscopy in the hard X-ray band ($>~6~\mathrm{keV}$) and the 
NuSTAR mission~\citep{2013ApJ...770..103H}, which was meant primarily for
observations of other astrophysical sources but is capable of hard X-ray imaging 
spectroscopy ($>~3~\mathrm{keV}$) of the Sun as well. Broad-band spectral measurements 
with these instruments allowed the solar X-ray spectrum to be modeled over a wide range of energies 
to probe various aspects, including the contribution of the non-thermal processes in the corona to the X-ray emission. 
However, as the lower energy threshold of these instruments is higher than $1~\mathrm{keV}$, there are limitations in constraining the thermal component in the emission, particularly during low solar activity.
It may also be noted that RHESSI has completed its mission duration and sensitive solar observations 
with NuSTAR are feasible only under quiescent solar conditions. 

In the absence of instruments providing imaging and broad-band spectroscopy 
extending down to the energy of $1~\mathrm{keV}$, instruments that carry out even 
spatially integrated measurements, like those on board various planetary 
missions, are of importance.
Such measurements over a wide energy range in soft X-rays have been carried out sporadically
over the past two decades by 
a few dedicated experiments: \textit{Solar X-ray Spectrometer}
(SOXS) on board GSAT-2~\citep{2005SoPh..227...89J},
\textit{Solar Photometer in X-rays} (SphinX) on board CORONAS-Photon mission~\citep{2013SoPh..283..631G}, 
and the recent \textit{Miniature X-ray Solar Spectrometer} 
(MinXSS) CubeSat missions~\citep{2018SoPh..293...21M}.  
Given the lack of dedicated solar instruments providing broad band spectroscopy, 
such measurements carried out by the Chandrayaan-2 XSM 
aptly complement the observations from wide-band X-ray imagers, narrow-band spectrometers, 
and hard X-ray spectrometers for investigations of the solar corona.

The XSM on board the Chandrayaan-2 mission provides disk integrated 
solar spectra in the energy range of 1 -- 15 $\mathrm{keV}$ with a spectral resolution 
of better than 180 eV at 5.9 $\mathrm{keV}$, which is the best available so far 
among similar instruments that carried out such measurements. The XSM also offers 
the highest time cadence for such instruments: full spectrum every second 
and light curves in three energy bands every 100 ms.  
The unique design features of the XSM allows observations over a 
wide dynamic range of X-ray fluxes from the quiet Sun to X-class flares.  
Presently, the XSM is the only instrument operational providing soft X-ray 
spectral measurements of the Sun over a broad energy range.  

The Chandrayaan-2 spacecraft was launched by the Geosynchronous Satellite Launch Vehicle (GSLV) 
MkIII-M1, on 22 July 2019.
It reached its nominal circular orbit around the Moon in early
September after several orbit maneuvers and the XSM began its nominal operations 
in lunar orbit from 12 September 2019. In-flight observations with the XSM have 
been used to evaluate its performance and validate and refine the ground calibration. 
In this article, we present the onboard performance and calibration of the XSM and prospects of 
solar studies with the instrument. Section~\ref{xsm_inst_obs} provides an overview 
of the instrument, observation plan, and data analysis; the onboard performance of the instrument 
is discussed in Section~\ref{xsm_inflight}; and some of the science cases that can be addressed 
with the XSM are presented in Section~\ref{xsm_science}, followed by a summary.

\section{Chandrayaan-2 Solar X-ray Monitor: An Overview}
\label{xsm_inst_obs}

\subsection{Instrument and Ground Calibration}

The XSM is designed to carry out spectroscopic observations of the Sun in X-ray wavelengths with stable spectral 
performance over a wide range of solar X-ray intensities. As the objective is to
measure the disk integrated solar X-ray spectrum, the instrument does not
have any imaging elements. It uses a Silicon Drift Detector (SDD) to measure the 
energy of individual photons and records the X-ray spectra in the 1--15 $\mathrm{keV}$ energy range 
with a cadence of one second. The detector is covered with a detector cap or 
collimator having a small aperture such that the entrance area of the instrument 
is restricted while maintaining a large field of view (FOV) of $\pm$40 $\mathrm{degree}$.  
This large FOV maximizes the visibility of the Sun as the instrument is fix-mounted on 
the spacecraft and the angle between the Sun vector and the instrument boresight 
varies over a wide range depending 
on the attitude configuration of the spacecraft~\citep{vanitha20}. The choice of materials and design of the 
collimator ensures that the  background from all other directions and any fluorescence
emission is blocked, thereby ensuring that the XSM has a very low background~\citep{mithun20_gcal}.

The XSM employs a closed-loop control of the temperature of the detector to ensure 
that the spectral resolution of $\approx 175 ~\mathrm{eV}$ does not vary during in-flight 
observations where the ambient temperatures are expected to vary significantly. The unique characteristics 
of the detector and the readout system also make it possible to maintain this spectral performance, 
without significant effect of pulse pileup, up to an incident flux of 
about 80,000 $~\mathrm{counts~s^{-1}}$~\citep{mithun20_gcal}, 
which corresponds to the M5 class of flares as discussed later. To further extend the dynamic 
range, the XSM includes a filter wheel mechanism which brings a 250$~\mathrm{\upmu m}$ beryllium 
window in front of the detector when the count rate exceeds a set threshold. 
The Be window attenuates X-rays below 2 $\mathrm{keV}$ thereby, reducing the count rate and thus enabling 
spectral measurements even for higher intensity flares. The filter wheel also includes 
an Fe-55 radioactive source covered with Ti foil, which serves the purpose of carrying out in-flight 
calibration of the spectrometer. 
A summary of the major specifications of the instrument is given in Table~\ref{xsm_specification} and
a detailed discussion of the instrument design is given in \cite{shanmugam20}.

 \begin{table}
 \caption{Specifications of the XSM.}
 \label{xsm_specification}
 \begin{tabular}{l l}
 \hline
 Parameter & Specification \\
 \hline
 Energy Range & 1 -- 15 $\mathrm{keV}$ (up to $\approx$ M5 class) \\
              & 2 -- 15 $\mathrm{keV}$ (above $\approx$ M5 class)\\
 Energy Resolution & $<$ 180 $\mathrm{eV}$ @ 5.9 $\mathrm{keV}$ \\
 Time cadence & 1 $\mathrm{s}$ \\
 Effective area (on-axis) & 0.135 $\mathrm{{mm}^2}$ @ 1 $\mathrm{keV}$ \\
                          & 0.367 $\mathrm{{mm}^2}$ @ 5 $\mathrm{keV}$ \\
 Field of view    & $\pm$40 degree \\
 Filter wheel mechanism properties & \\
 \hspace{0.5 cm}Positions & 3: Open, Be-filter, Cal(Fe-55)\\
 \hspace{0.5 cm}Be-filter movement threshold flux & 80,000 $\mathrm{counts~s^{-1}}$ ($\approx$ M5 class) \\
 \hline
 \end{tabular}
 \end{table}

 In order to infer the incident solar spectrum from XSM observations, the instrument's 
 spectral response needs to be calibrated. Several dedicated ground calibration experiments were carried out to determine 
 the gain parameters of the XSM under various observing conditions of ambient temperature and Sun angle, 
 the spectral redistribution function of the detector, and the effective area as a function of incident 
 angle. These measurements were used to derive an on-ground estimate of the response 
 matrix. A detailed description of ground calibration aspects of the XSM is presented in \cite{mithun20_gcal}.

\subsection{Estimation of Count Rates for Different Classes of Solar Flares}

\begin{figure}
 \centerline{\includegraphics[width=0.7\textwidth]{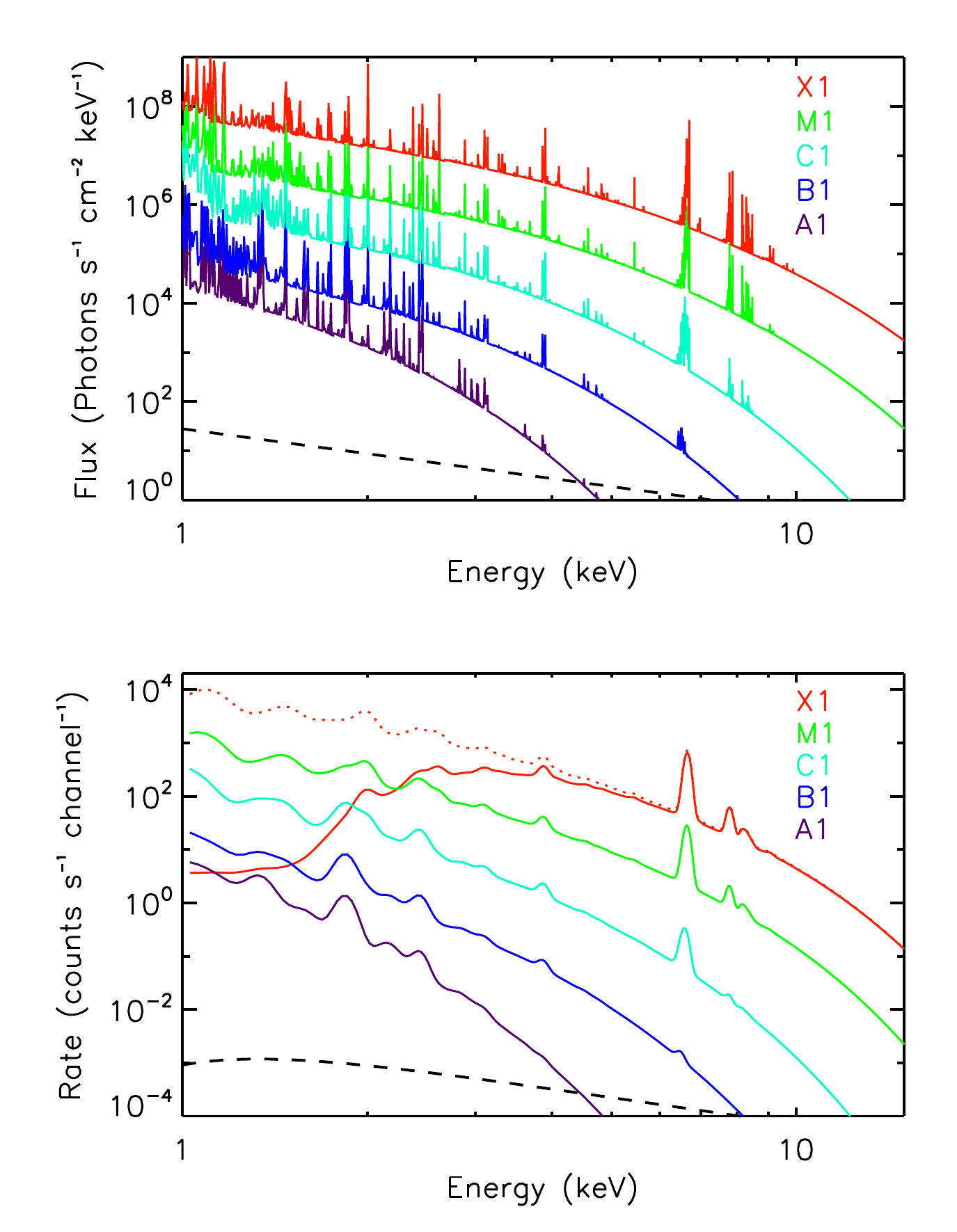}}
 \caption{Top: Model spectra for different flare classes computed using CHIANTI. Bottom: Model
  spectra convolved with the XSM spectral response showing expected observations. 
  Black dashed lines on both plots show the estimated Cosmic X-ray Background spectrum.}
 \label{simspec}
 \end{figure}

 We utilize the response of the XSM instrument obtained from ground calibration to estimate
 the expected count rates during solar observations at various levels of solar activity.
 For this purpose, the CHIANTI 
 atomic database~\citep{1997A&AS..125..149D,2015A&A...582A..56D} was used. 
 Synthetic solar spectra were generated considering isothermal plasma emission comprising of both 
 continuum and lines.
 This was done for a range of temperatures and emission measures that span different classes of solar 
 flares based on the correlations from ~\cite{1995ApJ...450..441F}. 
 The upper panel of Figure~\ref{simspec} shows such synthetic spectra for flare classes ranging from A1 to X1 
 while the lower panel shows the expected spectra from observations with the XSM obtained by convolving the model 
 spectra with the on-axis response. In the case of an X1 class flare, the dotted line shows 
 the spectrum without the beryllium filter and the solid line shows the spectrum
 with the beryllium filter placed in front of the detector thereby 
 increasing the low energy threshold to $\approx 2~\mathrm{keV}$.

 Since the XSM is an instrument with a wide FOV, the most important source of background is the
 diffuse Cosmic X-ray Background (CXB), which will determine the sensitivity of the 
 instrument. We estimate the expected background spectrum in the XSM using the CXB spectral
 model given by ~\cite{turler10}. The CXB spectrum per unit solid angle obtained from the model is 
 multiplied with the solid angle within the FOV of the XSM to obtain 
 the incident CXB spectrum, which is shown with a black dashed line in the 
 top panel of Figure~\ref{simspec}. This model spectrum
 is convolved with the XSM response matrix and is shown with a black dashed 
 line in the bottom panel of the same figure. The total count rate due to CXB is  
 0.1 $\mathrm{counts~s^{-1}}$.
 It is seen from the figure that the estimated background spectrum is about two orders
 of magnitude below that of the A1 class spectrum at lower energies. Hence, it is expected 
 that the XSM can provide spectral measurements, even when solar activity is below the A1 class. 

 Another possible contribution to the background in the XSM is from persistent 
 X-ray sources that are within the FOV of the instrument. 
 However, even for the Crab X-ray source, which is the standard candle in X-ray astronomy and one of 
 the brightest X-ray sources, the expected count rate in the XSM is 0.03 $\mathrm{counts~s^{-1}}$, which 
 is negligible in comparison to the CXB background. 
 Apart from the X-ray background, charged particles and secondary emission produced by their interaction 
 with satellite structures can also contribute to the background in the XSM. 
 However, as the XSM detector is protected with package walls from all directions except for 
 the small aperture, this contribution is expected to be of the 
 same order or lower than the CXB background. 

 \begin{figure}
 \centerline{\includegraphics[width=0.7\textwidth]{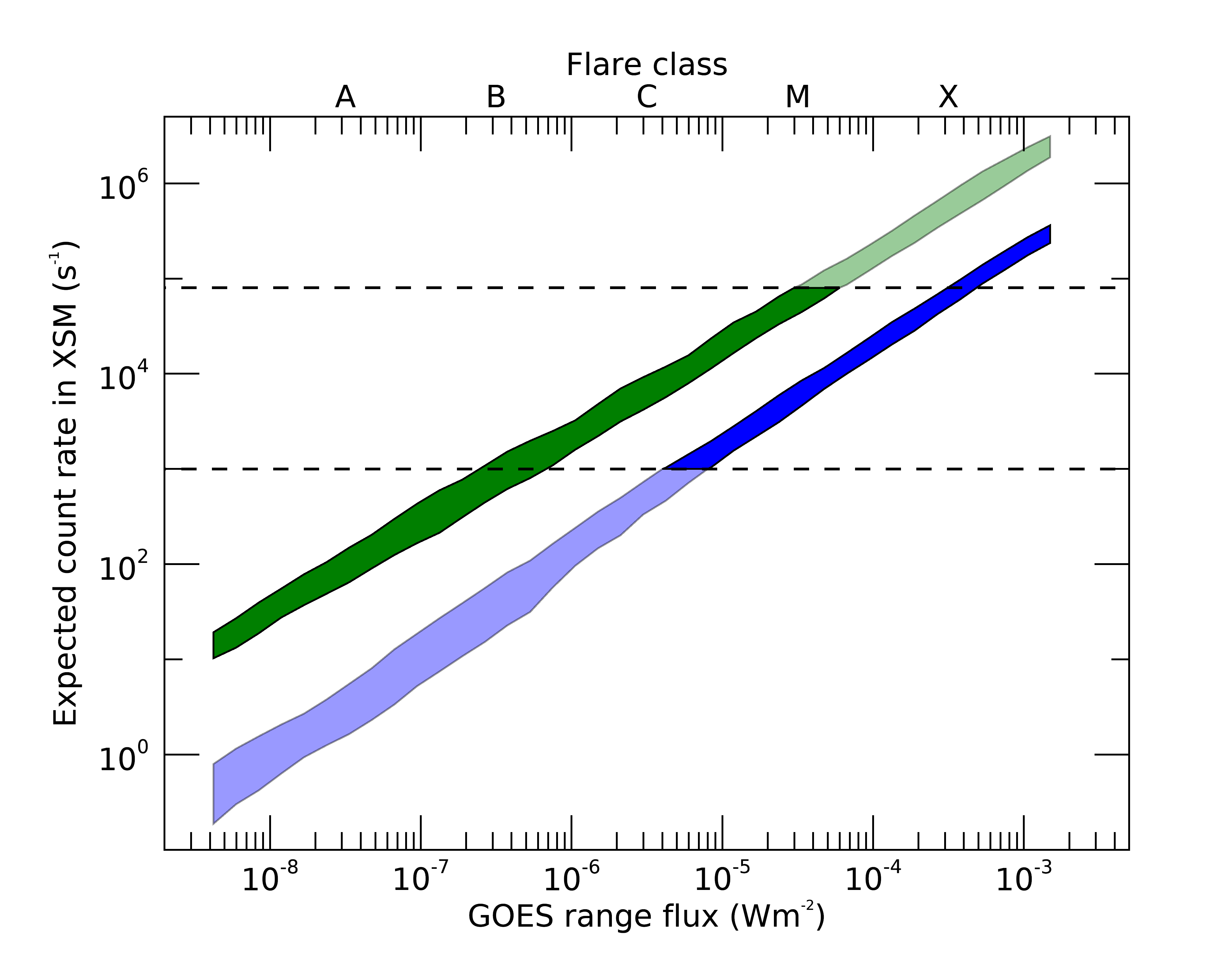}}
 \caption{Expected count rate in the XSM for different flare classes. Blue and green shaded 
 regions correspond to count rates with and without the Be filter, respectively. Dark colors 
 represent the expected modes of operation for the threshold rates for filter change shown 
 by the horizontal dashed lines. The XSM operates without the Be filter (shown in dark green) up to a count rate of 
 80,000 $\mathrm{counts~s^{-1}}$, beyond which it operates with 
 the Be filter in front of the detector (shown in dark blue) until the count rate drops below 
 1000 $\mathrm{counts~s^{-1}}$.} 
 \label{flare_crate}
 \end{figure}

 In order to determine the nominal range of solar flare classes where the XSM will operate with and 
 without the beryllium filter, the expected count rates are computed from the simulated spectra. 
 The estimated count rates are shown as a function of flare class in Figure~\ref{flare_crate}. 
 Shaded regions represent the range of count rates expected based on the typical 
 range of temperatures and emission measures for each  flare class presented in~\cite{1995ApJ...450..441F}.    
 The green shaded region shows count rates without the thick Be filter, whereas the blue 
 shaded region corresponds to rates when the Be filter is present in front of the detector.
 Considering that the transition to the Be filter occurs at 80,000 $\mathrm{counts~s^{-1}}$ and 
 the transition back to open position is at 1000 $\mathrm{counts~s^{-1}}$, the dark blue and 
 dark green regions in the figure represents the nominal operation modes with and without the Be window. 
 These simulations suggest that 
 the transition to the Be filter would occur around the M5 class of solar activity. Also, up to X5 class of solar activity, 
 the count rates stay within $10^5$ $\mathrm{counts~s^{-1}}$, where no degradation is expected in the spectral 
 performance~\citep{mithun20_gcal}. It may also be noted that above X5 and up to X9 class, though the spectral resolution 
 would be slightly inferior ($\approx 220~\mathrm{eV}$), the observations would still be useful. 

\subsection{Observation Plan and Data Analysis}
\label{xsm_data}

Solar observations with the XSM are planned according to the two orbital seasons of the Chandrayaan-2
spacecraft~\citep{mithun20_gcal}, each spanning approximately three months.
During the three months of the `dawn-dusk' season (D-D), 
the XSM observes the Sun almost continuously. At the beginning and the end of this season,
there are short durations ($\approx 20-30$ minutes per $\approx 120$ minute orbit) of occultation of the Sun by the Moon, 
which vanishes as the D-D day approaches. 
During the $\approx 30$ day period around the D-D day,
the XSM has almost uninterrupted observations of the Sun. 
Apart from the occultation period, there can be short periods ($\approx 10$ minutes per orbit, a couple
of times a day) of operation of other instruments on the spacecraft during which Sun is out of the field of
view of the XSM.
During the `noon-midnight' season (N-M), the XSM has a much lower cadence of observations.
In the initial and last $\approx 20$ days of the season, the attitude definition is such that
the Sun is completely out of the field of view of the XSM, and hence no observations are available
during this time. After the initial $\approx 20$ days, the Sun enters into the FOV of the XSM 
with an exposure time of few minutes that increases up to $\approx 25$ minutes per orbit 
until the N-M day. 
After the N-M day, exposure per orbit starts decreasing, and the Sun is not in FOV 
for the last $\approx 20$ days of the season. 
Table~\ref{season_def} gives the approximate days of the year covered by both 
the seasons. 

 \begin{table}
 \caption{Approximate duration of orbital seasons of Chandrayaan-2.}
 \label{season_def}
 \begin{tabular}{c c c}
 \hline
 Season & Period 1 & Period 2 \\
 \hline
 Dawn-Dusk & August 10 - November 20 &  February 14 - May 20 \\
 Noon-Midnight &  November 20 - February 14 &  May 20 - August 10 \\
 \hline
 \end{tabular}
 \end{table}

Data from all payloads of the Chandrayaan-2 spacecraft are downloaded at the Indian Deep
Space Network (IDSN) and the Deep Space Network (DSN) ground stations~\citep{vanitha20}. 
After pre-processing at the Indian Space Science Data Center (ISSDC), Bangalore, 
the XSM level-0 data sets are sent to the Payload Operations Center (POC) located 
at the Physical Research Laboratory (PRL), Ahmedabad, where the higher-level data 
processing is carried out. Raw (level-1) and calibrated (level-2) data sets of the XSM, 
organized into day-wise files, are then archived following the Planetary Data System-4 (PDS4) 
standards. The XSM data will be made available publicly from the ISRO Space Science 
Data Archive (ISDA) at ISSDC after a lock-in period of a maximum of nine months 
after each observing season.

For the analysis of the XSM data, a user-level software named 
XSM Data Analysis Software (XSMDAS) and the required calibration database (CALDB) 
will also be made available~\citep{mithun20_soft}. 
XSMDAS consists of individual modules to generate data products 
with a user-defined set of input parameters. All the data files of the XSM generated 
by XSMDAS are in FITS format. The raw data include the payload data frames, 
housekeeping information, and various observation geometry parameters. Using the 
XSMDAS modules light curves, spectra, and associated response matrices can be 
generated from the raw data with the required time or energy ranges and 
bins. The XSM data archive also contains standard calibrated products which are 
generated by the default processing at the POC. These include a time-series 
spectrum and total light curve for the full energy range, both with a time bin 
size of one second.  

Spectra and response files generated by XSMDAS are directly usable 
with standard spectral analysis tools used in X-ray astronomy such as XSPEC~\citep{arnaud96} 
and ISIS~\citep{2000ASPC..216..591H}. The time-series spectrum file which is 
part of the data archive can also be loaded into OSPEX, which is an IDL-based 
spectral fitting program in the SolarSoft (SSW) distribution. For this purpose, 
an IDL routine has been provided as part of the XSMDAS distribution. 
Detailed descriptions of the XSM Data Analysis Software and algorithms, 
data processing at the POC, and the contents of the XSM data archive are 
presented in \cite{mithun20_soft}.

\section{In-flight Performance and Calibration}
\label{xsm_inflight}

After the launch of Chandrayaan-2, the XSM was first powered-on for a short duration 
in earth-bound orbit to verify its performance. Similar short observations were 
carried out in the lunar orbiting phase to investigate any effect of passage 
through the earth's radiation belt. Regular observations with the XSM started on 12 September 2019, 
and it has been operating almost continuously since then.
The XSM acquires data even when the Sun is not within the FOV in order to obtain background measurements. 
The instrument is powered off occasionally for short periods during orbit maneuvers 
and other mission-critical operations.

The performance of the XSM has been evaluated during the six months of in-orbit operations 
with the onboard calibration source, background measurements, as well as solar observations,
as discussed in this section. We examine the stability of the energy resolution and gain 
using calibration source spectra and compare the performance with the ground measurements, and 
using both background and solar observations, we establish the sensitivity limit of the XSM.
Further, with a careful analysis of the quiet Sun observations at different Sun angles, 
we provide a refinement to the effective area which was obtained from ground calibration alone. 

\subsection{Energy Resolution and Gain}

\begin{figure}
 \centerline{\includegraphics[width=1.0\textwidth]{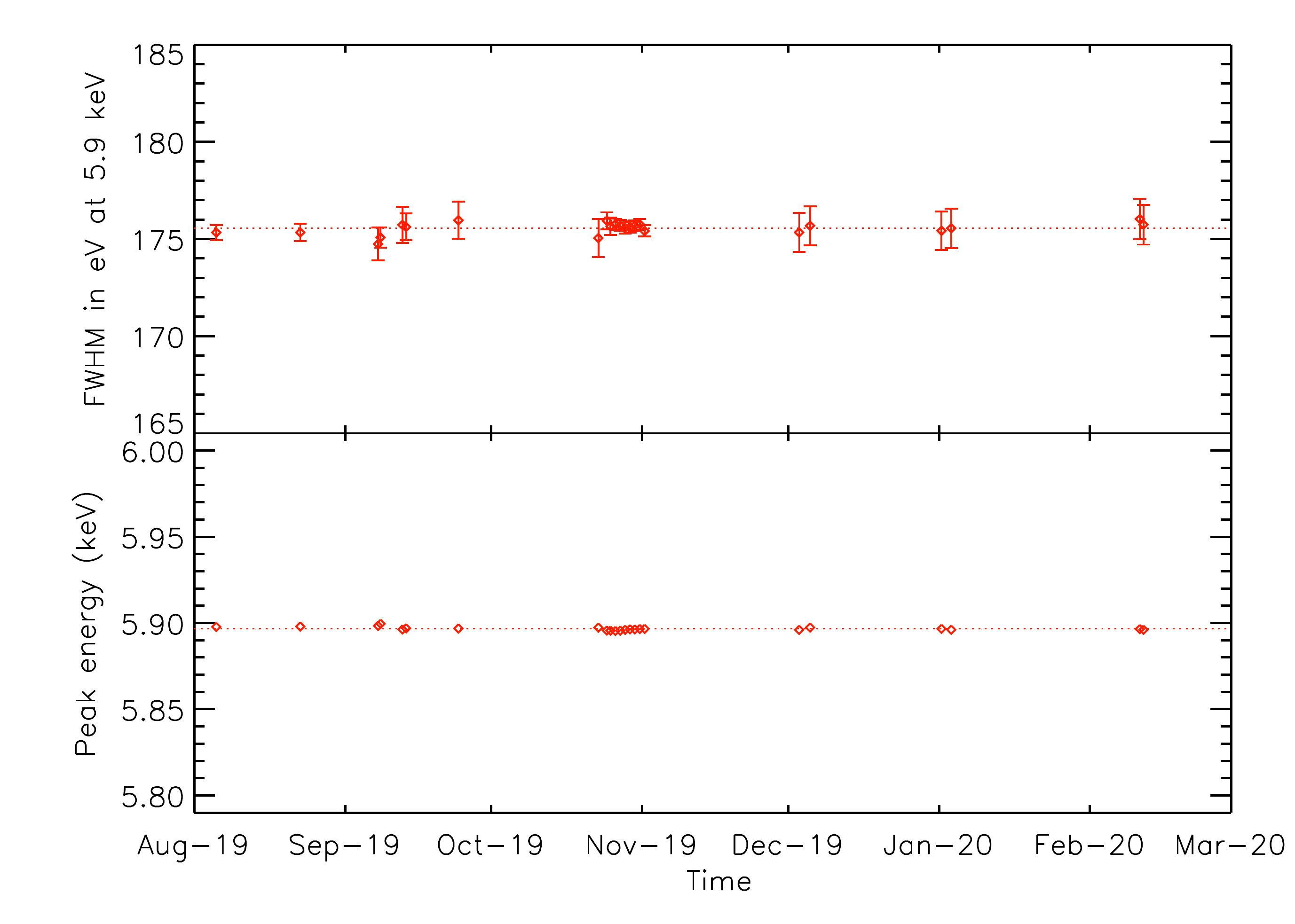}}
 \caption{The energy resolution (FWHM) and estimated peak energy for the 5.9 $\mathrm{keV}$ line 
 from the onboard calibration source for the first six months of the in-flight operation 
 of the XSM. Dotted lines show the mean values.}
 \label{inflight_fwhm_gain}
 \end{figure}

The spectral performance of the XSM is being monitored by using the onboard calibration source 
mounted on the filter wheel mechanism. 
Data were acquired with the calibration source during the in-orbit commissioning 
phase and at regular intervals since. The raw calibration spectra obtained were 
corrected for gain using the parameters obtained during ground calibration. 
Lines in the pulse invariant spectra thus obtained were fitted 
with Gaussians. The spectral resolution, defined by the full width at half maximum (FWHM) 
of the line at 5.9 $\mathrm{keV}$, 
over time is shown in Figure~\ref{inflight_fwhm_gain}. The peak energy of the 
same line is also shown as a function of time in the bottom panel of the 
figure. 

It can be seen that the spectral resolution remains the same as that obtained
prior to the launch 
and has remained constant over the six-month duration in-orbit.
This shows that the instrument is working flawlessly, and there is 
no degradation in the performance after the spacecraft's passage through the earth's radiation 
belts and due to the lunar environment. 
Line energy as estimated from gain corrected XSM spectra is consistent 
with the incident energy and shows no variation with time.
This confirms that the ground calibration obtained for the gain holds good in space, 
and there is no variation in the gain parameters so far. Monitoring of 
the gain and resolution of the XSM will continue with calibration source observations, 
and the CALDB will be updated in case of any changes in these 
parameters.

\subsection{Background and Sensitivity}

\begin{figure}
 \centerline{\includegraphics[width=0.9\textwidth]{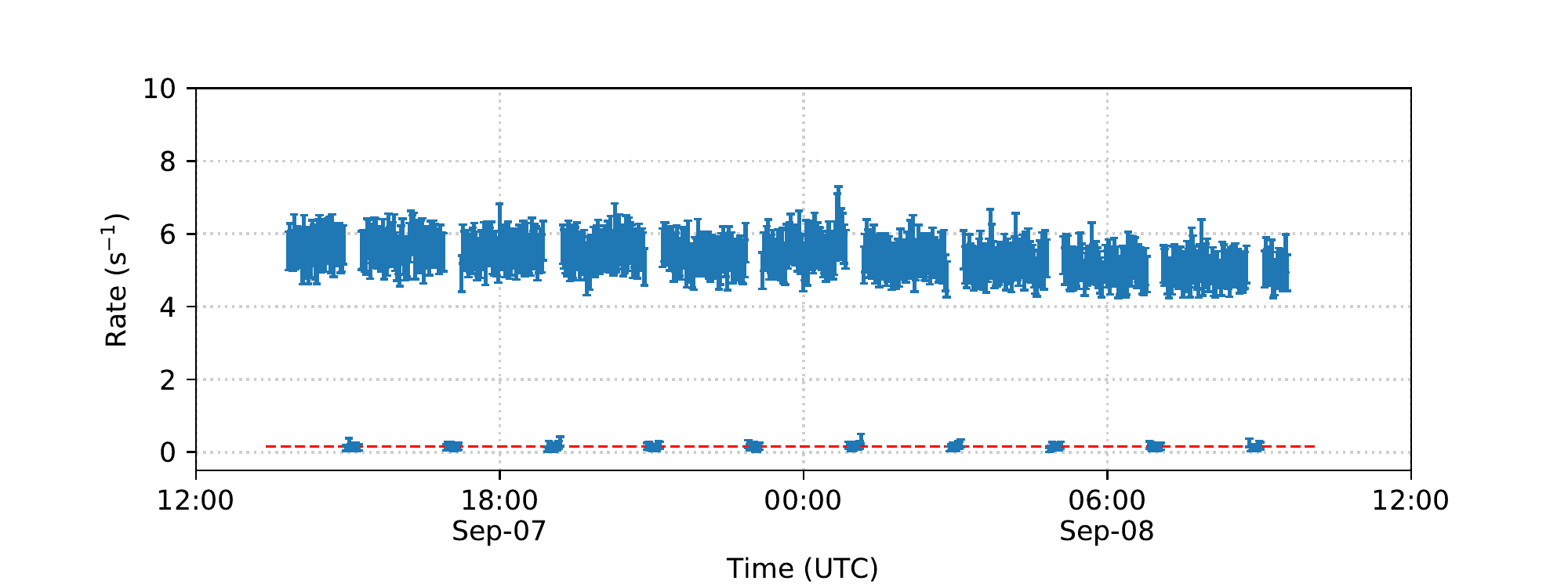}}
 \caption{XSM light curve with 100 s bin size during 07-08 September 2019 with periods of solar observations 
 and occultation by the Moon. The red dashed line shows the mean background count rate during the periods when the Sun 
 was occulted by the Moon, which is $\approx 0.15$ $\mathrm{counts~s^{-1}}$. 
 During this period of very low solar activity, counts from the Sun are detected by the XSM 
 well above background.}
 \label{inflight_bkg}
 \end{figure}

 The background rate in the XSM detector primarily determines the sensitivity for solar observations during 
 low solar activity periods. Background measurements are 
 available when the Sun is either occulted by the Moon or is out of the XSM field of view. In the commissioning 
 phase, on 07-08 September 2019, observations of the Sun with the XSM were carried out with 
 intervening durations of occultation. Figure~\ref{inflight_bkg} shows the count 
 rate observed with the XSM during this observation showing both the background and the quiet Sun. 
  The observed background rate is $\approx 0.15$ $\mathrm{counts~s^{-1}}$ very close to the estimated 
 CXB rate of 0.1 $\mathrm{counts~s^{-1}}$. The difference is attributed to the additional contribution 
 from particle-induced background.
 As seen from the figure, the background is approximately 35 times lower than the count 
 rate from the Sun even during this quiet period when the solar activity was well below 
 the A1 level. It may be noted that apart from the $\approx 0.15$ $\mathrm{counts~s^{-1}}$ background events, the XSM also records 
 $\approx 1-2$ $\mathrm{counts~s^{-1}}$ events that deposit energy greater than the high energy threshold of the XSM which are 
 recorded in the last channel of the instrument that is ignored for all spectral analysis. 
 These events are also due to high energy particle interactions in the 
 detector, but they do not affect the spectral measurements. 
    
 \begin{figure}
 \centerline{\includegraphics[width=0.9\textwidth]{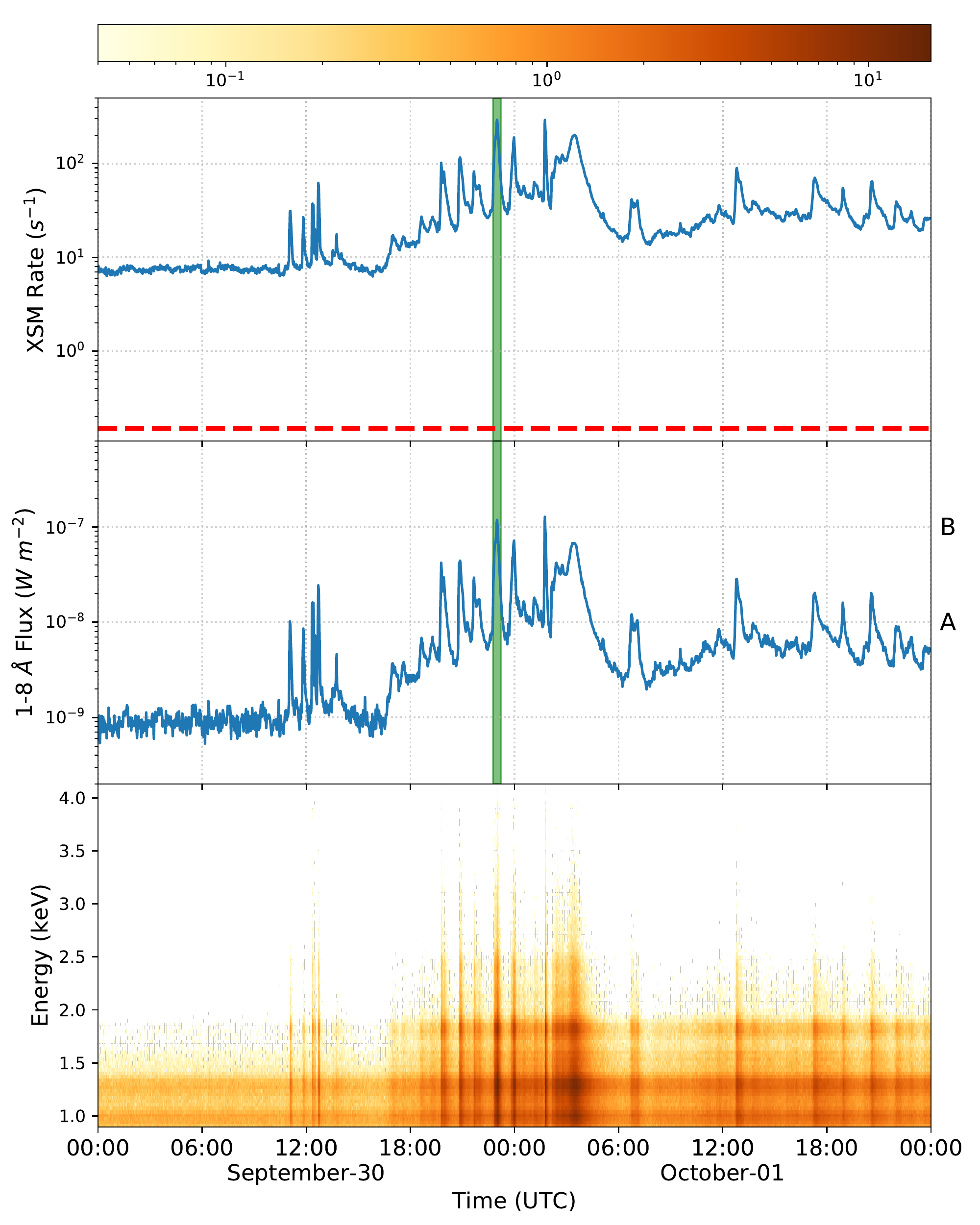}}
 \caption{Solar light curve (top), X-ray flux in the 1-8 $\mathrm{\AA}$ (1.55-12.4 $\mathrm{keV}$) range estimated 
 from the XSM data (middle), and  the dynamic spectrum (bottom) with a time bin of 100s. 
 In the top panel, the red dashed line corresponds to the background count rate.  The duration includes 
 periods of very low activity showing a few A-class and sub-A class flares, and two B-class flares. The dynamic spectrum 
 shows the spectral variability during the flares.  The integrated spectrum for the shaded duration around 
 the peak of the first B class flare is given in Figure~\ref{xsm_flare_spec}.}
 \label{xsm_flare}
 \end{figure}

\begin{figure}
 \centerline{\includegraphics[width=0.9\textwidth]{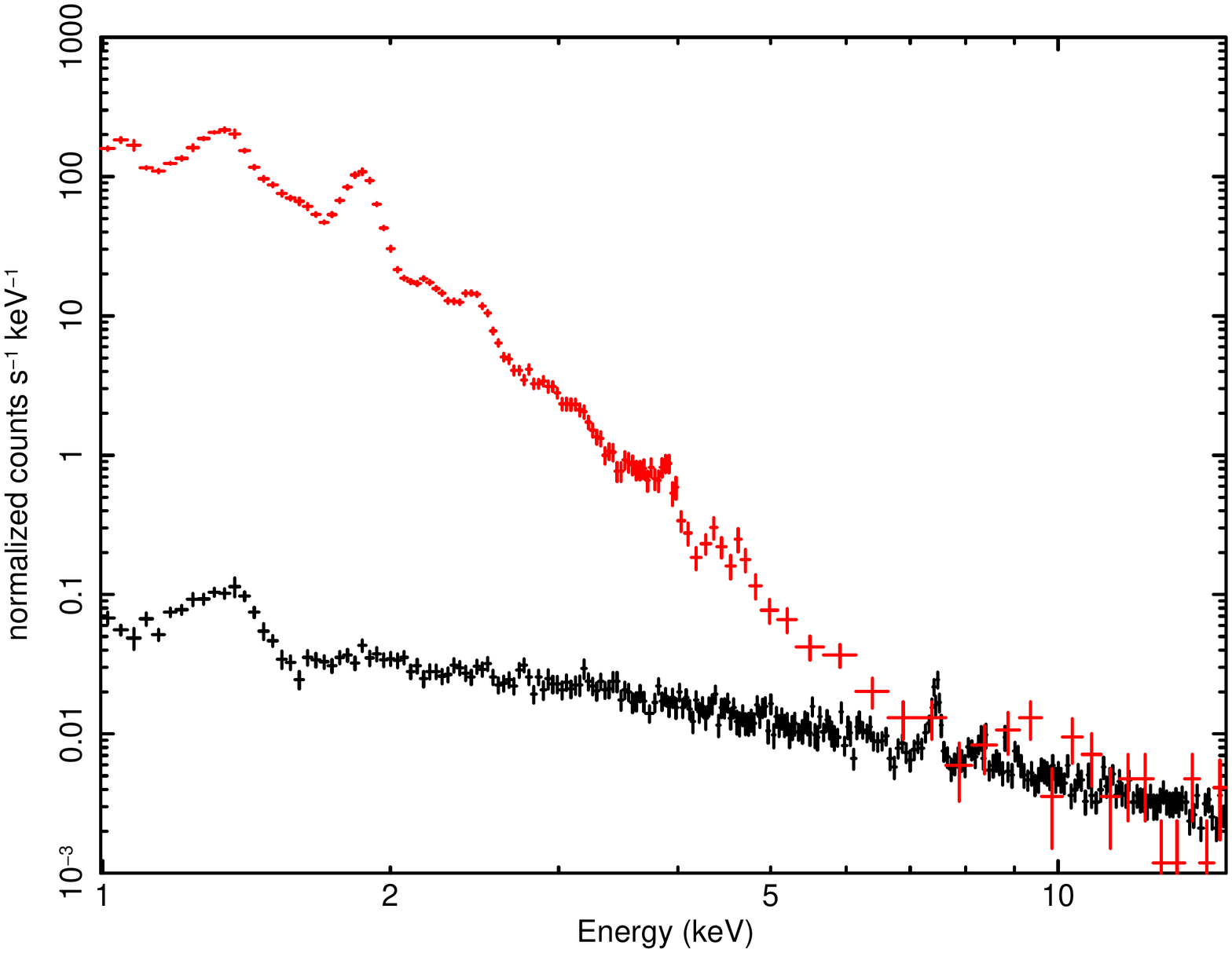}}
 \caption{Solar spectrum (red) as measured by the XSM during the B1 class flare on 30 September 2019 
 for the shaded duration (1700 $\mathrm{s}$) shown in Figure~\ref{xsm_flare} . A representative background spectrum (black) 
 of the XSM is also shown for comparison. The line seen in the background spectrum at $\approx 7.5~\mathrm{keV}$ 
 is of instrumental origin as mentioned in the text.} 
 \label{xsm_flare_spec}
 \end{figure}

 During September 30 - October 01 of 2019, two B1 class flares occurred that were also detected 
 by the GOES XRS instrument. Figure~\ref{xsm_flare} shows the XSM observations 
 during this period. The top panel shows the observed light curve in 100 second 
 bins, the middle panel shows the flux estimated for the 
 wavelength range of 1-8 $\mathrm{\AA}$ (the GOES range) by integrating the XSM spectrum within this range, 
 and the bottom panel shows the dynamic spectrum for the same duration. 
 For comparison, the typical background rate is shown in the top panel with a dashed line.
 It is evident from the plots that even when the solar activity was an order of magnitude below 
 A1 class, the count rates in the XSM, for the Sun, are significantly higher than the background. 
 Spectral variability during the flares is evident from the dynamic spectrum. 
 The integrated spectrum for the duration of the peak of the first B1-flare, marked by the 
 shaded region in Figure~\ref{xsm_flare}, is given in Figure~\ref{xsm_flare_spec}. For comparison, 
 the background spectrum is also shown. It is to be noted that for low-intensity B1 class flares,
 spectroscopy at low energies up to $\approx 6~\mathrm{keV}$ is unaffected by the background; 
 however, at higher energies, the background is important to consider. 
 It can be seen that the background spectrum
 has a line at $\approx 7.5~\mathrm{keV}$, which corresponds to Ni-K$\alpha$. This arises from
 a thin nickel layer present between the aluminium of the collimator and its silver coating, and the same 
 was observed during ground calibration as well. As the energies of the lines in the solar spectra 
 during large flares (see Figure~\ref{simspec}) do not coincide with that of this line, it is not expected to 
 interfere with the spectroscopic analysis.

\begin{figure}
 \centerline{\includegraphics[width=0.99\textwidth]{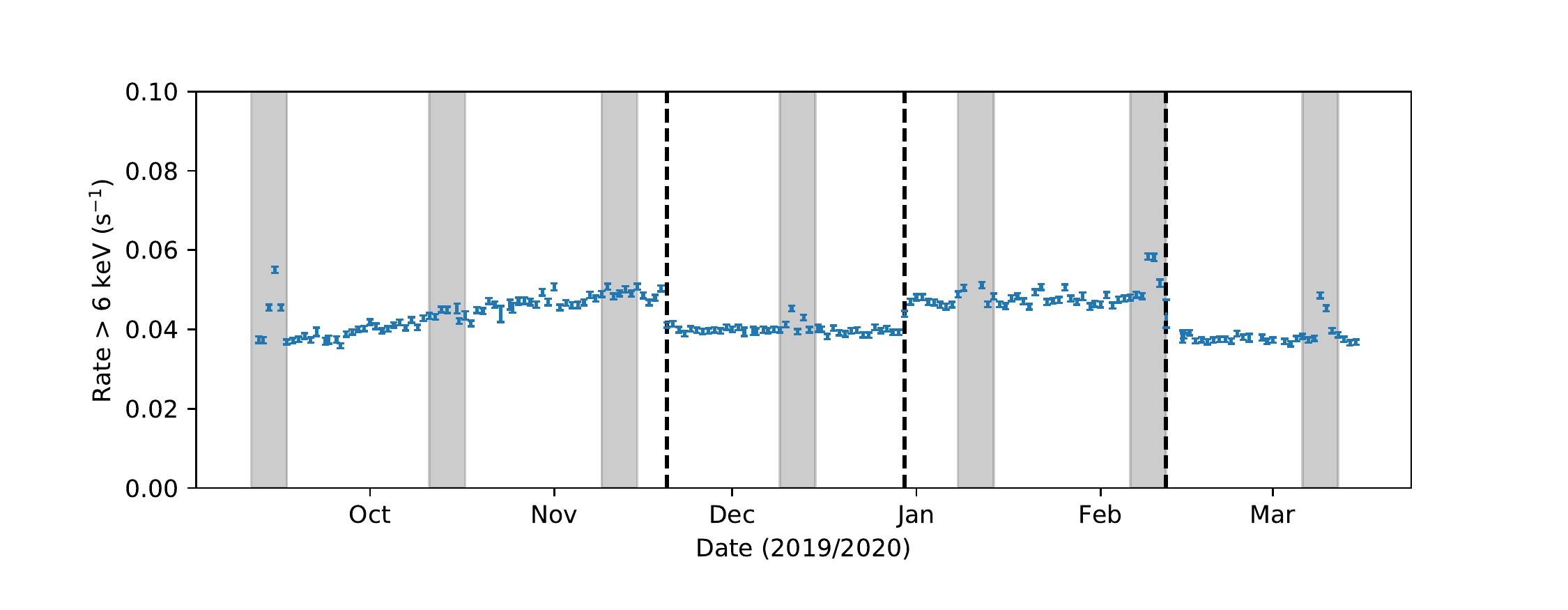}}
 \caption{The daily average XSM count rate above 6 $\mathrm{keV}$ plotted as a function of time. 
 The variability seen is associated with background variations. Vertical dashed lines 
 show the instances of change in attitude configuration of the spacecraft, which coincides 
 with sharp changes in the count rate.  Gray shaded regions correspond to the passage 
 of the spacecraft through the earth's magnetospheric tail, where 
 enhanced particle concentrations are expected.}
 \label{bkg_lc}
 \end{figure}

 Cosmic X-rays, the primary component contributing to the background in the XSM, are not expected to 
 vary with time. On the other hand, the particle-induced background can be variable. 
 In order to investigate the variation of the background in the XSM, we examine the light 
 curve in the higher energy band above 6 $\mathrm{keV}$ over the six months of in-flight operation. 
 As the solar activity was low during this period, at these energies, the contribution 
 from the Sun is expected to be negligible in comparison to the background. 
 Figure~\ref{bkg_lc} shows the XSM light curve above 6 $\mathrm{keV}$, where each point is 
 the mean value for the day. The light curve clearly shows a small but distinct variability in 
 the background. In the first part, there is a systematic increase in the daily average 
 count rate, which then has sudden variations in three later instances. The vertical dashed lines 
 in the figure represent the times when the attitude configuration of the spacecraft was 
 changed that coincide with the sudden changes in the count rates. 
 Hence, it is understood that there is a correlation between the attitude configuration 
 and the background in the XSM. There may also be further variations within a day that are 
 averaged out in this plot.
   There are also multiple short periods with enhanced background rates that coincide with the passage of the spacecraft 
 through the magnetospheric tail of earth, marked by the gray shaded regions in the figure.  
 An increase in the background in these durations is expected from the enhanced particle densities 
 in the geo-tail~\citep{2014LPI....45.2199N}.
 It may be noted that the magnitude of variations in the background is very small compared  
 to the solar flare spectrum, and hence it will not affect the spectral analysis of 
 solar flares, which is the primary goal of the XSM. However, in order to extend the spectral analysis 
 of the quiet Sun to higher energies, the background variations need to be understood and 
 modeled, which is planned to be carried out in the near future.

 Based on the in-flight observations of the background as well as the Sun, it can be seen 
 that the XSM is sensitive enough to measure solar activity in the soft X-ray band at 
 flux levels of at least two orders of magnitude lower than A1 class events. 
 The XSM is capable of detecting flare events that fall below A1 class and also 
 provides their spectral measurements.  The XSM has observed many such flares, and 
 the results will be reported in a future publication. 

\subsection{Collimator Response and Effective Area}
\label{effarea_inflight}

The effective area of the XSM is critically dependent on the collimator response and 
parameters, such as the thickness of the window, detector thickness, and dead layer. 
The collimator response was experimentally determined for estimating the effective area, 
whereas, for the other parameters, which are internal to the detector module, 
the manufacturer provided values were used~\citep{mithun20_gcal}.
Here, we examine the adequacy of the ground calibration estimate of the effective area 
using the in-flight observations. In this section, we discuss the investigation of 
whether the field of view of the XSM is symmetric as understood from the ground 
calibration experiment and the re-calibration of the effective area with observations 
of the quiet Sun.

\subsubsection{Field of View}

\begin{figure}
\centerline{\includegraphics[width=0.99\textwidth]{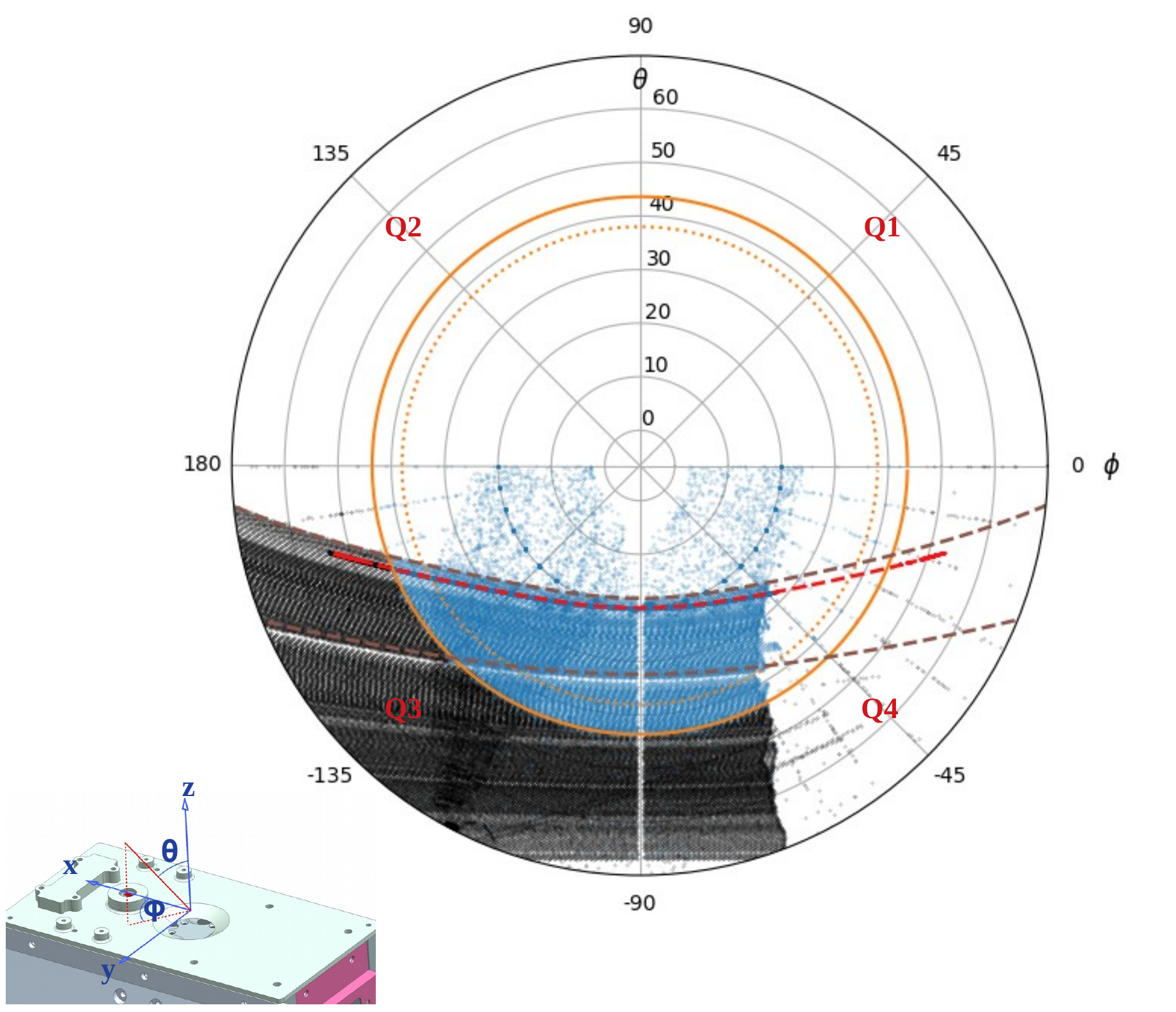}}
\caption{Polar plot showing the position of the Sun in the XSM reference frame and 
the corresponding count rate. Each point in the plot corresponds to the position of the Sun during a 
time bin of 10 s, and its color represents the count rate observed by the XSM.
Time bins with count rates at $5\sigma$ higher 
than the background rate are shown in blue, and others are in black.
The definition of the XSM reference frame is shown in the schematic at the 
bottom left corner of the figure. The center of the polar plot corresponds to 
the XSM boresight with polar angle $\theta$=0 and the radial axis represents 
$\theta$ values up to  $70^{\circ}$. The azimuthal angle ($\phi$) is defined in 
the range from $-180^{\circ}$ to $+180^{\circ}$.  
It may be noted that due to the mounting geometry of the XSM and the specific spacraft attitude 
configuration, all observations have $\phi$ within a range of $-180^{\circ}$ to $0^{\circ}$ 
(quadrants 3 and 4) and mostly have $\theta > 20^{\circ}$. 
The red dashed line shows the track of the Sun within the XSM FOV during nominal D-D season observations, and the brown dashed-lines show parts of the track during two representative days in the N-M season.
Durations of the occultation of the Sun, which happens only in quadrant 4,  
are not included in this plot, which results in the asymmetry between quadrants 3 and 4.
The solid orange circle represents the null points of the XSM FOV as obtained from 
ground calibration, and the dotted orange circle represents the full FOV. 
It can be seen that the blue points lie within the FOV, which shows that 
the FOV as determined from ground calibration is consistent with the onboard 
solar observations (see text for further details).}
\label{fov_result}
\end{figure}

When the Sun is at least partially within the field of view of the XSM, 
the recorded count rates are significantly higher than the background rate 
of $\approx 0.15$ $\mathrm{counts~s^{-1}}$. 
So, the XSM count rate can be used to verify whether the Sun is within 
the field of view or not, thus identifying the null points, i.e., the edge of the FOV of the XSM.
We used all available observations until March 
2020 for this purpose. Light curves with a bin size of 10 seconds were generated 
for the entire period, ignoring the durations when the Sun was occulted by the Moon.
For each time bin, average polar($\theta$) and azimuthal ($\phi$) angles of the Sun with respect to 
the XSM instrument frame were computed.  
Figure~\ref{fov_result} shows the polar plot of the position of the Sun during each time bin, 
where the radial axis represents the polar Sun angle. The time bins having 
count rates $5\sigma$ higher than the background rate are shown with blue points, 
and the others are shown with black points. The solid orange circle shows the edge of the FOV 
as obtained from the ground calibration experiment. 
It can be seen that the blue points 
representing the presence of Sun in the FOV are all within the orange circle, 
which shows that the onboard observations are consistent with a symmetrical field 
of view as measured on the ground.
It also demonstrates that the estimates of the Sun angle are correct.  
This is important to ensure because the Sun angle is used in estimating the 
effective area for a given observation.

\subsubsection{Effective Area Calibration}

Generally, X-ray spectrometers utilize observations of the standard source Crab
for in-flight calibration of effective area, but this is not feasible for the XSM
due to its very small aperture area and large field of view.
Hence, we have to rely on the quiescent Sun observations when
the inherent source variability is minimum. As the angle between the Sun
and the XSM bore-sight varies within each orbit during nominal observations,
that data can be used to characterize the effective area as a function of angle, 
avoiding any requirement of separate calibration observations.

\begin{figure}
\centerline{\includegraphics[width=0.9\textwidth]{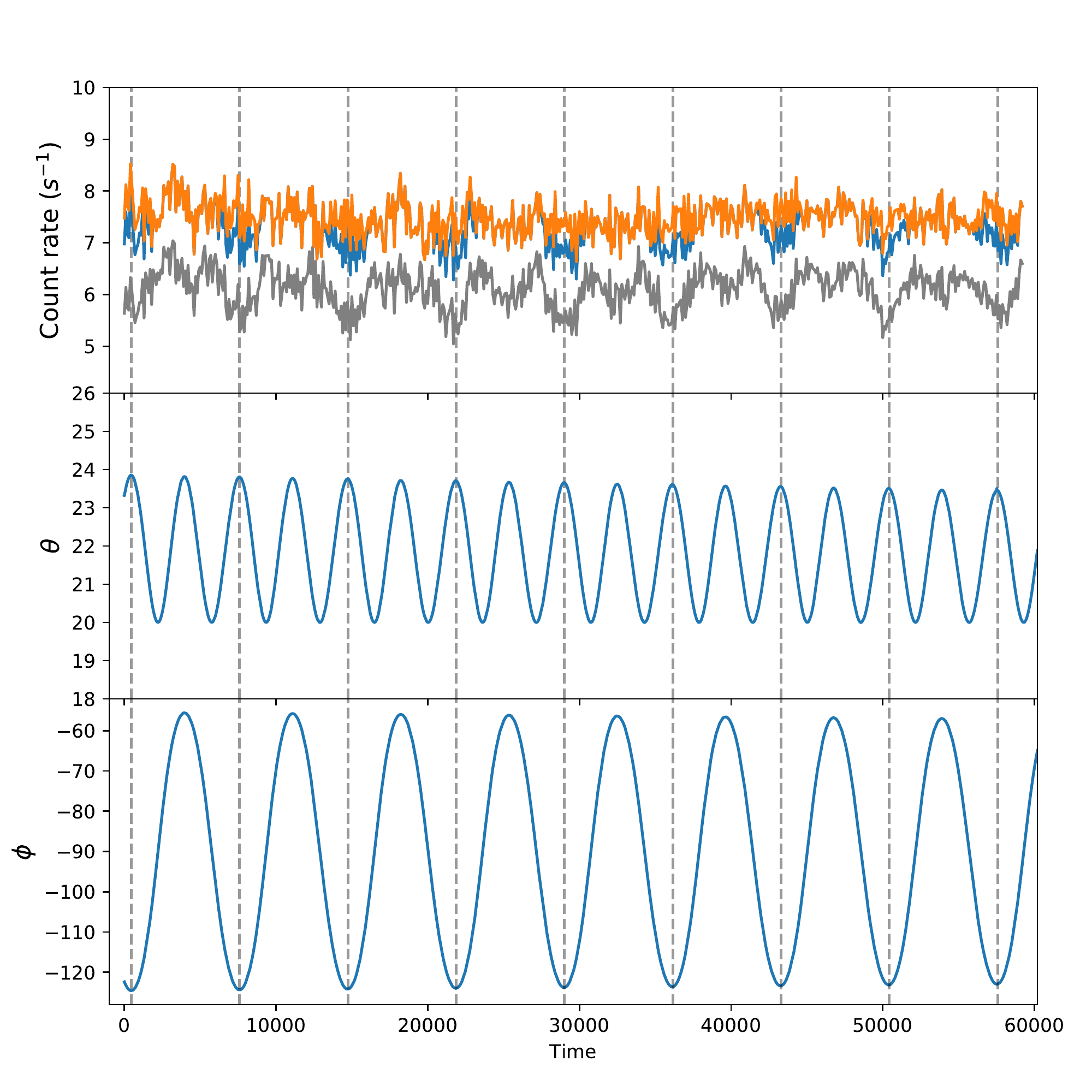}}
\caption{ Light curves obtained with the XSM on 17 September 2019 when the Sun was quiet 
without much inherent variability in the X-ray flux are shown in the top panel. 
The raw light curve (gray), the light curve corrected with ground calibration 
effective area (blue), and that corrected with the updated 
effective area (orange) are shown. 
Note that the mean value of the effective area corrected light curves are higher 
than the raw light curve as they are scaled to provide count rates corresponding to on-axis observations with the XSM.
The polar($\theta$) and azimuthal($\phi$) Sun angles are 
shown in the middle and bottom panels. The raw light curve shows significant 
modulation during the minima of the azimuthal angles marked by the vertical 
dashed lines in the figure, which has not been corrected by the effective area from 
ground calibration. With the refined effective area correction, the modulations in the 
light curve have been almost completely removed.}
\label{lc_angresp}
\end{figure}

As the angle between the Sun and the XSM boresight varies with time, the raw light 
curve shows modulation; however, these modulations are expected to get corrected 
after considering the corresponding effective area as determined from the ground calibration. 
The top panel of Figure~\ref{lc_angresp} shows 
the raw (gray) and effective area corrected (blue) light curves for the observation of 
on 17 September 2019 when the Sun was quiet and showing low X-ray variability. 
The other two panels of the figure show the polar ($\theta$) and azimuthal 
($\phi$) Sun angles during the same period.
It can be seen that even after incorporating the effective area correction 
based on the ground calibration, the modulation in the light curve  over 
the orbital phase where $\phi < -90^{\circ}$ is not fully corrected. 
A possible reason for this could be asymmetry in the FOV; however, this is ruled 
out based on the analysis presented in the previous section.
Thus, the uncorrected modulation over half of the orbital phase 
shows that the actual effective area has an azimuthal angle dependence and 
suggests that the effective area obtained from ground calibration 
needs further refinement. 

For this purpose, we considered all the available observations during the `dawn-dusk' 
seasons up to March 2020 and identified the durations when the variability in solar flux 
was minimum. This was done by examining the light curves manually and ignoring durations 
when even very small flares or any other variability within a day's timescale were present. 
Finally, observations from 42 days were 
selected for the analysis. It may be noted that for all these observations which 
are in the `dawn-dusk' season, the Sun follows the same track in the XSM reference frame, which is 
shown by the red dashed line in Figure~\ref{fov_result}.

\begin{figure}
\centerline{\includegraphics[width=0.75\textwidth]{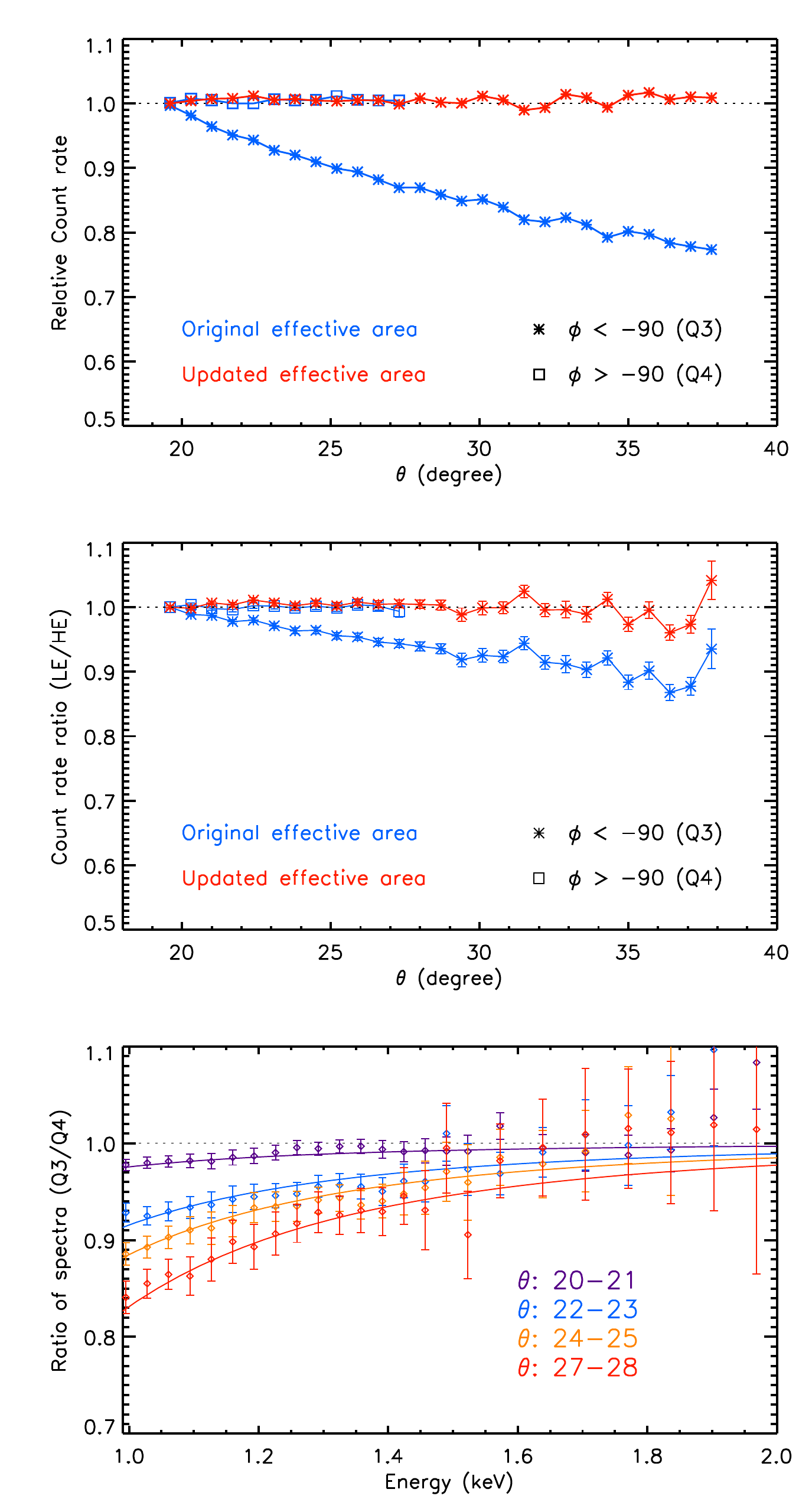}}
\caption{ Top panel: Mean XSM count rates during quiet Sun observations 
normalized to that at $\theta=20^{\circ}$ as a function of polar angle ($\theta$), 
for observations with $\phi<-90^{\circ}$ (Q3) and $\phi>-90^{\circ}$ (Q4).     
In the case of Q4, the count rates corrected with the original effective area (blue open square) remain constant with angle
confirming that the effective area is valid, whereas the count rates show a decreasing trend with 
angle for observations in Q3 (blue star).   
Middle panel: Normalized ratio of count rates in the energy ranges 1--1.2 $\mathrm{keV}$ (Low energy or LE) 
to 1.2--1.5 $\mathrm{keV}$ (High energy or HE)
as a function of $\theta$. The ratio shows a monotonic decrease with the angle in the case of Q3 (blue star), suggesting 
that the correction required in the effective area is energy dependent.  
Bottom panel: Ratio of spectra for observations in Q3 to that in Q4 for different bins of $\theta$. 
Solid lines show the additional absorption required for observations in Q3 
to explain the observed deviation of spectral ratio from unity at lower energies.
This correction factor is incorporated in the XSM effective area (for Q3), 
and the resultant corrected count rates and the ratio of count rates in two energy bands 
are shown with red points in the top and middle panels, respectively. The count rates 
and the ratio remains constant with $\theta$ which demonstrates the efficacy of the 
correction in the effective area}
\label{angresp_onboard}
\end{figure}

From data for each day, we generated light curves in the 1--15 $\mathrm{keV}$ 
energy range, scaled by the effective area from ground calibration, with a time bin size of 10 $\mathrm{s}$, and 
computed the corresponding mean polar ($\theta$) and azimuthal ($\phi$) 
Sun angles. 
To investigate the observed asymmetry in the effective correction with orbital phase, 
we estimated seperately, the mean count rate as a function of $\theta$ for
$\phi < -90^{\circ}$ (in quadrant 3 of the XSM reference frame, hereafter Q3) and
$\phi > -90^{\circ}$ (in quadrant 4 of the XSM reference frame, hereafter Q4) separately.
In order to remove any incident flux variations between different days of observation, 
we scaled the count rates for different $\theta$ with the count 
rate at $\theta=20^{\circ}$ for each day. 
These day-wise normalized count rates for each $\theta$ 
were averaged to obtain the results shown in Figure~\ref{angresp_onboard} 
top panel with blue star (Q3) and open square (Q4) data points.
It can be seen that for Q4, the normalized count rate, 
corrected with the effective area from ground calibration, remains constant with angle, 
whereas in Q3, it decreases with increasing angle. 
This shows that the estimate of the effective area reduction with $\theta$ 
from ground calibration is correct for Q4, but for Q3 it is steeper than the 
ground estimate. 

This enhanced reduction in effective area for observations in Q3 could be due to
a difference in collimator response, in which case it would be independent of energy, 
or due to difference in window transmission, in which case it would be energy dependent. 
In order to probe whether it is dependent on energy or not, we used light curves in two energy ranges:
1.0--1.2 $\mathrm{keV}$ and 1.2--1.5 $\mathrm{keV}$. 
From these light curves for each day of observation, 
count rates were estimated as a function of $\theta$ for both quadrants, and a ratio 
of rates in the lower energy band to the higher energy band was computed for each $\theta$. 
The count rate ratio as a function of angle for each day was normalized to that 
at $\theta = 20^{\circ}$ and were averaged to obtain the result shown in the middle 
panel of Figure~\ref{angresp_onboard}. If the reduction in the effective area were due to 
an energy-independent factor, this ratio is expected to remain constant with angle. 
It can be seen from the figure that for Q3, the ratio of count rates does not remain 
constant with angle, suggesting that the reduction in effective area from ground calibration 
estimate is energy dependent. As the ratio is less than unity and decreases with the increase 
in angle, the effective area for Q3 seems to require an additional absorption factor that 
depends on angle.  

To quantify this additional energy and angle dependent absorption factor in effective area, 
we used the available spectral information. From the same data selected for the earlier exercise, spectra were generated 
in each one degree bin of $\theta$ for the two quadrants separately. Using them, 
ratios of spectra when the Sun is in Q3 to that in Q4 were computed for 
each $\theta$ bin for all days. These day-wise spectral ratios were averaged to obtain 
the final set of ratios. As observations in Q4 are limited to 
a $\theta$ of $28^{\circ}$, the ratios are available for $\theta$ bins up to $28^{\circ}$ 
only. A representative set of spectral ratios are shown in Figure~\ref{angresp_onboard} 
bottom panel. It is evident from the figure that fewer counts are 
detected at lower energies for observations in Q3 compared to Q4, and this 
deviation increases with angle, confirming the presence of an additional absorption 
in this quadrant.     

The exact origin of such absorption in only one region of the field of view is 
not yet clear. It probably arises from the detector module, maybe due to the varying 
thickness of the Be window.
However, it is difficult to derive any conclusions based on the available data. 
Hence, we determine this additional absorption factor for Q3 empirically from the observed 
spectral ratios by modeling it as absorption by beryllium. Solid lines overplotted on the observed 
spectral ratios in the bottom panel of Figure~\ref{angresp_onboard} shows the 
empirically determined effective area correction terms that correspond to absorption by 
additional beryllium whose thickness ranges from $\approx 0.2 \mathrm{\upmu m}$ to $\approx 1.6 \mathrm{\upmu m}$ for 
different angles.
It can be seen that the derived effective area correction matches the observations very well. 
As the spectral ratios are available only up to $28^{\circ}$, the correction 
terms for the effective area at higher angles were obtained by extrapolating the 
dependence on $\theta$. We then incorporate this effective area correction term 
in CALDB and generate light curves corrected for the updated effective area. 
Count rates and the ratio of count rates in two energy bands as a function of 
$\theta$ after scaling with this updated effective area are shown in red color on the 
top and middle panels of the Figure~\ref{angresp_onboard}, respectively. 
The count rate and the ratio are now constant with the polar angle showing the 
efficacy of the effective correction term obtained. The light curve with the updated 
effective area for the data of 17 September examined earlier and shown in Figure~\ref{lc_angresp} 
also demonstrates the same. 

Apart from the beryllium window thickness, another factor that would cause uncertainties
in the effective area is the thickness of the silicon dead layer in front of the detector.
The dead layer affects the effective area the most at energies just above the Si K edge around
1.84 $\mathrm{keV}$, which in turn, may cause uncertainties in the measurement of solar abundances
of Si.
It can be seen from Figure~\ref{angresp_onboard} that area
uncorrected spectral ratios do not have any significant deviations from unity at those energies, and hence it can be concluded that the dead layer thickness is not varying significantly over
the detector area. However, the absolute value provided by the manufacturer (100$~\mathrm{{\upmu} m}$ Si
and 80 $~\mathrm{{\upmu} m}$ SiO$_2$) may have uncertainties. To estimate its effect on
the Si abundance measurements, we simulated spectra assuming different dead layer thicknesses
ranging from zero to double the expected value. These were further fitted with the standard
response with the manufacturer provided values of dead layer thickness to obtain measurements
of Si abundances and all other parameters. We find that, even with the assumption
of extreme ranges of dead layer thickness, the abundances measured for Si are within $\pm 0.5\%$ of the
actual value, which is very small compared to the typical statistical errors
associated with abundance measurements from the XSM spectra. Further, as the effective 
area near line complexes of other elements are significantly less affected by the dead layer thickness, 
their abundance measurements will be unaffected. 
Hence, we conclude that the effect of
uncertainties on the dead layer thickness can be safely ignored.

In order to estimate the overall uncertainties in the relative effective area with the polar angle, 
the area corrected count rates in different angle bins shown in the top panel of 
Figure~\ref{angresp_onboard} were used. The standard deviation of the count rates 
is $\approx 0.8\%$ and hence we quote a conservative limit of 1\% uncertainty in the 
relative effective area with angle.
It may be noted that the present analysis was carried out with D-D 
observations, and thus we have a fairly robust estimate of 
the effective area for these ranges of $\theta$ and $\phi$. 
For the N-M case, this would serve as a starting point, but we plan to continue the 
investigation of $\phi$ dependence of effective area as a function of energy
as more data are acquired in this attitude configuration.
Further, during the XSM operation so far, the Sun has been extremely quiet with 
emission well below A-class for most of the time, thereby yielding very low 
count rates mostly limited up to 2 $\mathrm{keV}$.
These observations were used for the present analysis, and the absorption factors 
were estimated with the spectral ratios in the 1--2 $\mathrm{keV}$ band. Although these 
observations seem to suggest that the ratios flatten above $\approx 1.6~\mathrm{keV}$ or so, 
this is limited by statistics. Observations with higher count rate extending to 
higher energies, when the Sun becomes more active, 
are likely to provide further insights into this.
Another point to note is that the requirement of additional correction factor in the 
effective area concern only the observations when the Sun in Q3. 
In Q4, the effective area is well understood, and the in-flight observations are
consistent with ground calibration estimates. 
So, it would be good to verify the results for observations in both quadrants 
separately whenever feasible.
In any case, we plan to continue the investigations as and when additional 
observations become available and the updated effective area will be made available in 
the revised calibration database.

\subsubsection{Low Energy Threshold}

The low energy threshold below which the XSM does not detect photons is determined by a 
threshold pulse height setting in the readout electronics, which can be changed by ground
command. This ensures that the instrument does not record events due to very small
fluctuations present in the signal. After the initial commissioning phase,
the default value of the XSM threshold was set to a value that nominally corresponds to
$\approx 900 ~\mathrm{eV}$. However, the spectrum would not have a sharp cut-off at this energy; rather,
the detection probability starts to increase from zero at energies below $900 ~\mathrm{eV}$ 
and then reaches unity at some energy higher than $900 ~\mathrm{eV}$.
An error function can generally describe this feature near the low energy
threshold (e.g., \citealp{2016SPIE.9905E..1IP}).

On careful analysis of the solar spectra, after incorporating the corrections for the effective
area as described in the previous section, it has been noticed that the effect of the
low energy threshold extends up to about $1.30~\mathrm{keV}$. Hence, the predicted
counts from response in channels up to this energy would be higher than the actual observation. After the completion of
the second D-D observing season, the low energy threshold setting was optimized to bring
the effects of the electronic threshold to energies lower than $1~\mathrm{keV}$ 
while avoiding any
spurious events, and the default setting was changed from June 2020. Thus, for observations
before June 2020, it is recommended that the spectrum below $1.3~\mathrm{keV}$ is ignored
during fitting.
\newline

 \begin{figure}
 \centerline{\includegraphics[width=0.9\textwidth]{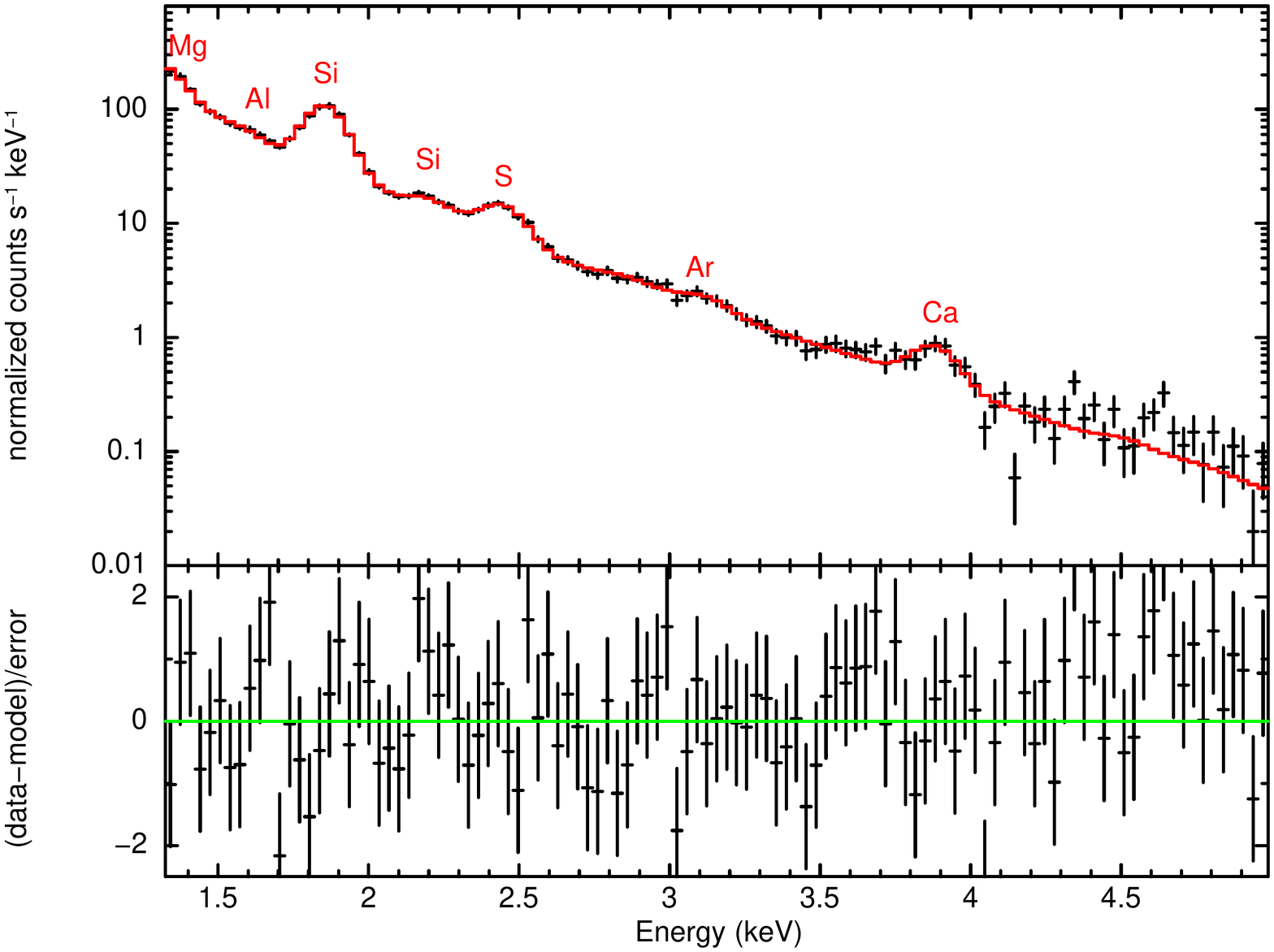}}
 \caption{Solar spectrum as measured by the XSM during the B1 class flare on 30 September 2019 shown in 
 the shaded time interval in Figure~\ref{xsm_flare_spec} fitted with \texttt{vapec} model in XSPEC. Spectrum in the energy range 
of 1.3 -- 5.0 $\mathrm{keV}$ is considered for the fitting. The best fit model for a temperature of $6.45~\mathrm{MK}$ is shown in red and the residuals are shown in the lower panel. Elemental line features are marked in the spectrum.}
 \label{xsm_fitspec}
 \end{figure}

To demonstrate the adequacy of the calibration status of the XSM, a representative flare spectrum on 30 September 2019 (same as in Figure~\ref{xsm_flare_spec}) was fitted with a model for
isothermal plasma emission. The model in XSPEC named \texttt{vapec}, which computes the
X-ray spectrum with the AtomDB atomic database, was used for this purpose. Figure~\ref{xsm_fitspec} shows
the spectrum, fitted model, and the residuals, where the best fit model corresponds to
a plasma temperature of $6.45 ~\mathrm{MK}$.
It can be seen that the model convolved with the XSM response fits the observed data very well.
Detailed spectral analysis and interpretations of solar observations with the XSM making use of an XSPEC model 
that uses the CHIANTI atomic database will be presented in future.

The XSM has been in orbit and operational for nearly one year at the time of writing. 
Its spectral performance has undergone no variation during this time and 
matches very well with the ground calibration. Gain corrections, taking into 
account the variations with electronics temperature as well as the X-ray 
interaction position in the detector, provides energy measurements within 
an accuracy of 10 eV. The observed background count rates are well 
within the expectations, and hence the XSM is sensitive enough to measure variability 
in X-ray flux even at the sub-A class level of solar activity. The effective area 
as a function of angle obtained from ground calibration is refined with 
the analysis of in-flight data, and the resultant effective area is shown 
to have an uncertainty of less than 1\%; hence the  error in flux measurements 
with the XSM will be of the same order. We plan to validate this further using 
more observations. We will also continue to monitor the calibration parameters of the XSM 
like gain and spectral resolution for any changes.

\section{Science Prospects}
\label{xsm_science}

The primary objective of the XSM is to aid the measurement of lunar surface elemental abundances
with the CLASS instrument. This requires observations during periods of increased solar
activity such that fluorescence emission from the lunar surface is significant. A preliminary result
of elemental abundance estimates during one low-intensity flare is presented in ~\cite{shyama20}.
Observations over the entire mission life are expected to provide abundance measurements
over substantial regions of the lunar surface.

Apart from this, the broadband X-ray spectral observations with the XSM with moderately high resolution
and unprecedented time cadence will enable one to explore specific science problems in solar physics.
Some of the areas where observations with the XSM can contribute are discussed below.

\subsection{Microflares}

Recent observations of the Sun with sensitive instruments like NuSTAR~\citep{wright_17} and 
FOXSI-2~\citep{2020ApJ...891...78A} have generated considerable interest in microflares. 
With its high sensitivity, NuSTAR has been able to detect the weakest  
(thermal energy of $1.1 \times 10^{26} \mathrm{erg}$) active region microflare so far~\citep{2020ApJ...893L..40C}.
Energetically, events classified as microflares are about a million 
times weaker events than standard flares. \cite{shimizu_95} had done the first number 
distribution study of such events using the Yohkoh Soft X-ray telescope and had found that 
a single frequency distribution with a power-law index ($\alpha$) between $1.5$ and 
$1.6$ can explain the occurrence frequency of flares and microflares together, 
up to an energy of $10^{27} \mathrm{erg}$. 
It is believed that such micro-flares, or nanoflares with even lower total energy, 
can deposit a significant amount of energy in the solar corona. However, in order to 
explain the coronal heating entirely as due to the micro/nano flares, it is 
required to have a steeper power-law index at lower energies~\citep{hudson_91}. 

As shown in the previous section, the XSM is quite capable of detecting sub-A class 
flare events having energies of the order of $10^{27} \mathrm{erg}$ and lower. 

As the XSM unlike NuSTAR and FOXSI, lacks imaging capabilities the intensity of faintest 
detectable flares are limited to the background quiescent coronal emission. However, the availability of 
continuous observations over long periods of time and the lower energy threshold extending down to 
$1~\mathrm{keV}$ make the XSM observations very useful and valuable. Particularly since the XSM is observing during the solar 
minimum conditions it would be feasible to detect fainter events and provide 
better constraints on the power-law index for the frequency distribution of microflares.    
Such measurements will help us in estimating the 
nature of the frequency distribution of even weaker nanoflares, 
and thus may be able to shed light on their contribution to the coronal heating. 

Microflares are often termed as Active Region Transient Brightenings (ARTB) \citep{girjesh_18}. 
During their evolution phase, they are often seen in ultraviolet bands, although 
they are undetectable in GOES~\citep{2019A&A...628A.134M}. 
Simultaneously observing such events with the 
XSM along with the Atmospheric Imaging Assembly (AIA) on board 
the Solar Dynamics Observatory (SDO) will help us to understand their multithermal nature. 
The SphinX X-ray spectrometer that operated for about nine months in 2009 detected several 
such events but did not have the UV context images from the SDO available as will be the 
case of observations with the XSM.
However, since the XSM observes the Sun as a star, one needs to make sure that at any instant, 
there has to be only one such event on the solar disk.  

\subsection{Non-thermal Emission in Flares and the Quiet Sun}

Non-thermal processes are dominant during the evolution of solar flares and
contribute significantly to the X-ray emission \citep{benz_05, 2008A&ARv..16..155K,2017SoPh..292..100D}.
A comparison of the observed X-ray spectrum with the model thermal spectrum
convolved with the instrument response would indicate whether a 
non-thermal component is present (e.g., \citealp{2013ApJ...771....1J,2014ApJ...791...23K}).
A time-resolved analysis of XSM spectra during flares would provide insights
into the evolution of the non-thermal component. 
Because of its better sensitivity and much higher spectral resolution 
compared to previous instruments, measurements with the XSM would especially be
useful in probing non-thermal emission during low-intensity flares where the
transition from thermal to non-thermal may occur at much lower energies ($\leq 10~\mathrm{keV}$).
Joint observations using the XSM and hard X-ray solar spectrometers such as STIX~\citep{krucker20} on board 
the recent Solar Orbiter mission and HEL1OS~\citep{sankar_17} on board the upcoming 
Indian mission Aditya-L1 will also be very useful. 
Joint fits to the spectra over a wider energy band combining the XSM and STIX/HEL1OS data will allow 
one to constrain the thermal and non-thermal components better.

In many cases, eruptive flares exhibit pre-flare activity, which is characterized 
in terms of the duration and intensity of soft X-ray flux prior to the impulsive rise 
of the flare emission~\citep{1996SoPh..165..169F,2007A&A...472..967C,2011ApJ...743..195J,2019ApJ...884...46M}. 
The X-ray flare light curves indicate that the pre-flare activity comprises of single 
or closely spaced multiple episodes of energy release (e.g., \citealp{2016ApJ...832..130J,2020SoPh..295...29M}).
Despite observational limitations, RHESSI observations have detected weak yet clear non-thermal 
component during the pre-flare phase (e.g., \citealp{2019ApJ...874..122H,2020ApJ...897..157S}) 
which points toward the role of small-scale magnetic reconnection in destabilizing 
the magnetic field configuration of the active region and thereby triggering 
subsequent eruptive flares. Such case studies are very limited in number 
and, therefore, statistically viable results cannot be inferred from them. Given 
the better spectral resolution in the energy band of 1--15 $\mathrm{keV}$, XSM data would provide 
an excellent opportunity to explore the pre-flare activity. 
The synthesis of XSM measurements with imaging data at 
multi-wavelengths (such as EUV, UV, and H$\alpha$) during the pre-flare activity would 
shed light on the triggering mechanisms of solar eruptions.

Non-thermal emission from the quiet Sun corona is considered as an indicator 
of the presence of nanoflare heated plasma. A careful 
spectral analysis of the XSM observations of the quiet Sun integrated 
over long periods can be used to probe this. As the XSM has observed the Sun 
during the solar minima, there are significant durations without any active 
regions present on the Sun allowing the spectral study of the quiescent corona 
even with disk integrated spectra.
Of course, such an analysis would require a better understanding of 
the background spectrum.

\subsection{Overall Temperature Structure of the Quiet and Active Corona}

As different regions of the solar corona demonstrate different temperature structure, 
it is customary to determine the Differential Emission Measure (DEM) of the whole-disk 
or that of a specific region, such as the active region or the 
coronal hole~\citep{landi_08, dudik_14, schonfeld_17}. The DEM is derived 
from the observed intensities and theoretical emissivities of spectral lines.  
For the quiet Sun, such DEM is seen to peak around log T = 6.0, while for
active regions, the DEM peaks around log T = 6.2 \citep{brosius_96}. 
Usually, while generating the DEM, most of the lines used are from Extreme Ultraviolet (EUV),
and hence, high-temperature regions of the DEM curves are not well constrained.
Observations and numerical simulations show that nanoflares can heat the local solar
plasma to temperatures of 10 $\mathrm{MK}$~\citep{2009ApJ...698..756R,2018arXiv180700763A}, suggesting that
the DEM curves of a nanoflare heated region should extend to similar temperatures.
To understand the slope of the DEM curve at this high-temperature, DEM analysis of
spectra at X-ray energies is required, which can be attempted using XSM observations.
However, in contrast to the EUV observations, DEM analysis with broad-band spectrometers
like the XSM would have to take into account the intensity in each energy band due to 
both line and continuum emission.

\subsection{Understanding Elemental Abundances of the Solar Corona}

It is well known that elements with low ($<10~\mathrm{eV}$) first ionization potential (FIP) are
generally at least twice more abundant in the corona than in the photosphere. This phenomenon
is known as the FIP effect ~\citep{2015LRSP...12....2L,2018LRSP...15....5D}. 
Low FIP elements are generally ionized in the chromosphere, while the high FIP elements 
remain at least partially neutral. Several models invoke various mechanisms such as force 
due to magneto-hydrodynamic waves for separation of ions from neutrals, 
to explain the observed enhancement of low FIP elements (see \citealp{2015LRSP...12....2L} for a recent review). 
During flaring activity, abundances in the corona are also found to change abruptly.
X-ray spectroscopy of the Sun can help us understand such behavior of elemental abundances more accurately.
There have been spectroscopic observations which suggest that the mean abundances of 
some of the elements reach photospheric 
values during flares~\citep{2014ApJ...787..122S,2015ApJ...805...49S}.
With high time cadence observations, it would be possible to understand the time evolution 
of abundances during flaring activity.
Broadband X-ray spectral measurements with the XSM in the 1 -- 15 $\mathrm{keV}$ range covers multiple strong lines
of many of the elements present in the corona. By fitting the observed spectra with theoretical models which makes use of atomic databases like 
CHIANTI, abundances of these elements can be estimated. 
It is expected that observations with the XSM can constrain the abundances of some of the elements in the 
quiet corona and active regions and also their evolution during different classes of 
solar flares.

\subsection{Long Term X-ray Flux Monitoring}

Solar X-ray flux has been continuously monitored by the GOES series of satellites over the past 
four decades. X-ray flux measurements from GOES have been used as indicators of the  solar 
activity over its eleven-year cycles (e.g., \citealp{2004SoPh..219..343J,2015A&A...582A...4J}), apart from other measurements, 
such as sunspot number and magnetic field strength. In recent times, it has been 
shown that the solar activity is declining based on a steady decrease in solar 
photospheric fields starting from around 1995~\citep{2015JGRA..120.5306J} and several 
other signatures~\citep{2011GeoRL..3820108J,2018A&A...618A.148J,2019SoPh..294..123S,2019SoPh..294...54S}. 
Consequently, the solar minimum in 2009 was unusually deep, as noted by many authors (e.g., ~\citealp{2011GeoRL..38.6701S}). 
During this period, on several occasions, the solar X-ray flux went below the sensitivity limit of GOES, 
thereby making accurate flux measurements with GOES not feasible. The SphinX spectrometer with better sensitivity 
that operated during this time recorded X-ray fluxes that were an order of magnitude below the 
GOES A1 class~\citep{2019SoPh..294..176S}. 

The minimum between Solar Cycles 24 and 25 in 2019-20 has been deeper than the previous one, 
and it is expected that the period of low activity may continue 
longer and may result in a weaker maximum compared to earlier cycles~\citep{bisoi_20a,2020SoPh..295...79B} .
It has been even speculated that the Sun may be headed for 
an extended period of very low solar activity similar to that experienced in the Maunder 
minimum~\citep{2015SunGe..10..147J}. If this trend continues, there is a need for more sensitive 
instruments to monitor the solar X-ray flux.
As discussed in the previous section, owing to the very low background, the XSM is sensitive 
enough to measure X-ray fluxes down to a few orders of magnitude below A1 class, and hence 
would be ideal to monitor the solar X-ray flux during such 
extremely low levels of solar activity and deep minima. Except for 
 $\approx 40$ days during each of the two `noon-midnight' seasons every year, observations with the XSM 
are available every day. Beyond the XSM, data from the X-ray spectrometers~\citep{sankar_17} and in-situ 
particle measurements~\citep{janardhan_17} on board the Aditya-L1 mission will enable one to 
continue the study of the Sun in this unique low activity phase.

\section{Summary}
\label{summary}

The \textit{Solar X-ray Monitor} on board the Chandrayaan-2 orbiter has been operational in
the lunar orbit from September 2019, and it is the only spectrometer currently monitoring
the Sun in soft X-rays. Observations of the Sun with the XSM have been continuing as per the
plan, and the instrument performance has remained stable.
Using the in-flight observations of the onboard calibration source, we have shown that the
gain and spectral resolution ($\approx 175~\mathrm{eV}$)  are stable over time
and they match the ground calibration.
The on-ground estimate of the effective area has been refined using in-flight observations, and the uncertainty in the effective area with angle is shown to be within 1\%.
In-flight observations have also demonstrated that the XSM has the sensitivity to carry out
spectral measurements during solar activity well below A class level, which is made possible
by the extremely low background.

With these established capabilities, the XSM is expected to contribute to our understanding of the Sun,
particularly with the investigations of the sub-A class flares, which are well detected
during the current low solar activity period. Although the nominal mission life for Chandrayaan-2 is two years, it is expected
that the operations would continue beyond that, which would then provide a unique
opportunity to monitor evolution the solar activity during at least the rising phase of the Solar
Cycle 25. With the commencement of the nominal operation phase of instruments on board the Solar Orbiter in next year,
there will also be opportunities for simultaneous observations with them.  Further, these
studies can also be continued with the X-ray spectrometers and other
instruments of the upcoming Aditya-L1 mission.

%

 \begin{acks}
We thank the reviewer for helpful comments and suggestions which helped in improving 
the presentation of the article.
The XSM payload was designed and developed by the Physical Research Laboratory (PRL), Ahmedabad, 
supported by the Department of Space, Govt. of India. 
PRL was also responsible for the development of
the data processing software, the overall payload operations, and data analysis
of the XSM. The filter wheel mechanism for the XSM was provided by the U. R. Rao
Satellite Centre (URSC), Bengaluru, along with the Laboratory for Electro-Optics 
Systems (LEOS), Bengaluru. Thermal design and analysis of
the XSM packages were carried out by URSC whereas, the Space Application
Centre (SAC) supported in mechanical design and analysis. SAC
also supported in the fabrication of the flight model of the payload and
its test and evaluation for the flight use. We thank various
facilities and the technical teams of all the above centers for their support
during the design, fabrication, and testing of this payload. 
The authors also thank the Chandrayaan-2 mission, operations, and ground segment teams.
We thank the CLASS team for useful discussions. The authors acknowledge G. Del-Zanna for 
his help with the simulations of theoretical spectra using CHIANTI and for useful discussions. 
The Chandrayaan-2 mission is funded and managed by the Indian Space Research Organisation (ISRO).
 \end{acks}



\begin{acks}[Disclosure of Potential Conflicts of Interest]

The authors declare that they have no conflicts of interest.

\end{acks}


%

\bibliographystyle{spr-mp-sola}
\bibliography{references_xsm.bib}  

\end{article}
\end{document}